\documentstyle[12pt]{article}

\newcommand{\beq}{\begin{equation}}
\newcommand{\eeq}{\end{equation}}
\newcommand{\beqa}{\begin{eqnarray}}
\newcommand{\eeqa}{\end{eqnarray}}
\newcommand{\beqar}{\begin{eqnarray*}}
\newcommand{\eeqar}{\end{eqnarray*}}

\newcommand{\ga}{\gamma}
\newcommand{\Ga}{\Gamma}

\newcommand{\inn}{\!\cdot\!}

\newcommand{\z}{\zeta}

\newcommand{\eg}{{\it e.g.,}\ }
\newcommand{\ie}{{\it i.e.,}\ }
\newcommand{\labell}[1]{\label{#1}} 
\newcommand{\reef}[1]{(\ref{#1})}
\newcommand\prt{\partial}
\newcommand\veps{\varepsilon}

\newcommand\bz{\bar{z}}
\newcommand\bF{\bar{F}}
\newcommand\bC{\bar{C}}

\newcommand\hV{\hat{V}}
\newcommand\hG{\hat{G}}

\newcommand\hp{\hat{\phi}}

\newcommand\hpsi{\hat{\psi}}

\newcommand\hS{\hat{S}}
\newcommand\hX{\hat{X}}

\newcommand\Tr{{\rm Tr}}

\begin{document}

\thispagestyle{empty} \rightline{\small  \hfill IPM/P-2003/053}
\vspace*{1cm}

\begin{center}
{\bf \Large S-matrix elements and  covariant \\
 tachyon action  in type 0 theory \\

 }
\vspace*{1cm}

{Mohammad R. Garousi}\\
\vspace*{0.2cm}
{ Department of Physics, Ferdowsi university, P.O. Box 1436, Mashhad, Iran}\\
\vspace*{0.1cm}
and\\
{ Institute for Studies in Theoretical Physics and Mathematics
IPM} \\
{P.O. Box 19395-5531, Tehran, Iran}\\
\vspace*{0.4cm}

\vspace{2cm} ABSTRACT
\end{center}

We evaluate the sphere level S-matrix element of two tachyons and
two massless NS states, the S-matrix element of four tachyons, and
the S-matrix element of two tachyons and two Ramon-Ramond vertex
operators, in type 0 theory. We then find an expansion for theses
amplitudes that their leading order terms correspond to a
covariant tachyon action. To the order considered,  there are no
$T^4$, $T^2(\prt T)^2$, $T^2H^2$, nor $T^2R$ tachyon couplings,
whereas, the tachyon couplings  $F\bF T$ and $T^2F^2$ are
non-zero.  \vfill \setcounter{page}{0} \setcounter{footnote}{0}
\newpage

\section{Introduction} \label{intro}
The discovery in the early days of string theory that string
amplitudes of massless states in low energy regimes may be
reproduced by Yang-Mills field theory for open strings and
gravitational field theory for closed strings, was the beginning
of a long, fruitful study into the relation of string theory and
field theory in general and in low energy in particular
\cite{anjs}.

Recent days, effective actions that include massless as well as
tachyon states playing an important  role in understanding the
dynamics of  tachyon condensation in D-branes of bosonic string
theory  and in non-BPS D-branes of superstring theory
\cite{mgas}-\cite{sen2}. Beside the S-matrix based approach
\cite{anjs}, there are other approaches to find these actions.
One is the method introduced in \cite{aatt} which is based on
derivative expansion of partition function \cite{ku,ms}. The other
one is based on directly integrating out the massive modes of
string field theory to find an  action that includes tachyon and
massless fields \cite{wt}. In general, however, the fields  in the
resulting effective actions are related to each other by some
field redefinition. For example, in the cubic string field
theory, after integrating out all the massive and tachyon fields,
one will find an effective action for D-branes which is not
Dirac-Born-Infeld action plus higher derivative terms \cite{wt}.
While the massless scalar fields have clear interpretation in the
cubic string field theory, they have no (geometrical)
interpretation in the effective action. However, using nontrivial
field redefinition, one will find a field variable in terms of
which the effective action rearranges into the DBI action plus
higher derivative terms \cite{wt}. The new scalar variables have
now the geometrical interpretation as the transverse coordinate of
the D-brane in the effective theory, whereas, they  have no clear
interpretation in the original string field theory. Since the
mass squared of tachyon and massive modes are of the same order,
it is not clear how to extend the work in \cite{wt} to find an
action which includes massless fields as well as the tachyon.

It has been speculated in \cite{mrga} that imposing the
non-abelian gauge symmetry on the tachyon action, the  S-matrix
based approach \cite{anjs} may be used to find  the tachyon-gauge
field action. In particular, it has been  suggested in \cite{mrga}
that there is a unique expansion for the S-matrix elements of
tachyon vertex operators  that their leading order terms are
consistent with the standard  non-abelian kinetic term. The
S-matrix element of four open string tachyons, and the S-matrix
element of two tachyons and two gauge fields have been analyzed
in \cite{mrga} in favor of the proposal. It has been found that
while the leading order massless/tachyonic poles in the expansion
are consistent with non-abelian kinetic term the next leading
contact terms, in the superstring theory, are in fact consistent
with the non-abelian extension of the tachyon Born-Infeld action
\cite{mg,ebmr}.

There are two problems  in the above discussion. While the action
consistent with the expansion of the S-matrix elements has the
non-abelian gauge symmetry manifestly,  the physical
interpretation of the expansion  is not clear. A different
expansion for the S-matrix elements in terms of spatial momenta of
the external states has been suggested in \cite{ku}. This
expansion has the physical interpretation as expansion around a
marginal state, however, it is not clear that the action
consistent with the leading terms of the expansion  have
manifestly non-abelian gauge symmetry. One may expect that the
action in the two case would be related to each other by some
field redefinition. The second problem is that it does not
indicate that the resulting action is an effective action. In
fact if one applies the above idea for first excited massive
scalars of the D-branes, one will find a similar result
\cite{mrga}. However, if one proves by other means that there is
a tachyon-gauge field effective action with non-abelian gauge
symmetry, then the above action should be the effective action,
because there is only one expansion for the S-matrix elements
that corresponds to non-abelian gauge symmetry.

In the present paper, we would like to apply the above idea to the
closed string tachyon of type 0 theories. That is, we would like
to find an expansion for the  S-matrix element of closed string
tachyon and massless states that their leading order terms
correspond to tachyon action with covariant symmetry. We will
call this expansion the covariant expansion.  If one prove by
other means that there is an effective tachyon-massless fields
action, then the action should be the effective action.

Spectrum of type 0 theories can be obtained by a diagonal GSO
projection on the superstring spectrum or by orbifolding the
corresponding type II theories by $(-1)^{F_s}$, the total target
space fermion number \cite{ldjh}. They are represented as \beqa
{\rm type}\,
0A&:&(NS_-,NS_-)\oplus(NS_+,NS_+)\oplus(R_+,R_-)\oplus(R_-,R_+)\,\,,\nonumber\\
{\rm type}\,
0B&:&(NS_-,NS_-)\oplus(NS_+,NS_+)\oplus(R_+,R_+)\oplus(R_-,R_-)\,\,,\nonumber\eeqa
which then consists, at lowest level, of tachyon and, at massless
level, of graviton, dilaton, Kalb-Ramond antisymmetric tensor, and
two  RR states with opposite chirality. As we will see later, an
essential part in finding the covariant expansion of  a S-matrix
element of tachyons is to  compare it with the corresponding
S-matrix element of a massless scalar state that its covariant
expansion is trivial. Hence, we compactify theory on a torus and
consider the scalar components of the dimensional reduction of the
graviton as the scalars in the massless level. Furthermore, we
will assume all closed string states to be independent of the
compact directions. This makes it easier to compare  the S-matrix
element of tachyon with the S-matrix element of the massless
states to find the covariant expansion of the tachyon S-matrix
elements.

The paper is organized as follows. In the following section we
calculate the S-matrix element of two massless NS and two
tachyons. In this case it is easy to  find the covariant
expansion  even without comparing it with the corresponding scalar
amplitude.  We then show that a covariant action of order
$\alpha'^2$ for the tachyon reproduces exactly the leading terms
of the covariant expansion of the  amplitude. This calculation
indicates that there is no $T^2H^2$ nor $T^2R$ couplings at this
order in the action. In section 3, we repeat the same calculation
for the S-matrix element of four tachyon states. The leading
terms of the expansion are again fully consistent with the
covariant tachyon action. This calculation  predicts the tachyon
potential $V(T)$ has no $T^4$ term, and the tachyon action has no
coupling $T^2(\prt T)^2$. In section 4, we do the above
calculation for two RR and two tachyon states. The details
analysis of  the first leading term in the covariant expansion of
the amplitude fixes the couplings $F\bF T$ and $F^2T^2$. Section
5 is devoted to the discussion and comments on our results. In
the Appendix A, we give the result for the integrals that appear
in evaluating the above S-matrix elements. In Appendix B, we
evaluate the S-matrix element of four RR states with opposite
chirality.

Before continuing with our calculations, let us make a comment on
conventions.  The spacetime is assumed to be orthogonal product of
compact torus and non-compact flat space. The non-compact
directions are labeled by $a,b,c,\cdots$ and compact directions
are labeled by $i,j,k,\cdots$. The closed string states are
assumed to have momentum only in the flat directions.  The
graviton, Kalb-Ramond  and RR polarizations are in the non-compact
directions, and the polarization of massless scalars are  in the
compact direction. Our conventions also set $\alpha'=2$.

\section{Two tachyon-two graviton amplitude}
In this section we analysis in details the S-matrix element of two
tachyons and two massless NS states. In string theory side this
amplitude is given by the following correlation function: \beqa
A(NS,NS,T,T)&\sim& \langle:V_{(-1,-1)}^{\rm
NS}(p_1,\veps_1):V_{(-1,-1)}^{\rm NS}(p_2,\veps_2):V_{(0,0)}^{\rm
T}(p_3):V_{(0,0)}^{\rm T}
(p_4):\rangle\,\,,\labell{amp3}\nonumber\eeqa where the tachyon
and the NS vertex operators are given as \beqa V^{\rm
T}_{(0,0)}(p)&=&\int d^2z\,:ip\inn\psi(z)e^{ip\cdot X}:
ip\inn\hpsi(\bz)e^{ip\cdot \hX(\bz)}:\,\,,\nonumber\\
V^{\rm NS}_{(-1,-1)}(\veps,p)&\!\!\!\!=\!\!\!\!&\veps_{ab}\int
d^2z:e^{-\phi(z)}\psi^a(z)e^{ip\cdot
X(z)}:e^{-\hp(\bz)}\hpsi^b(\bz)e^{ip\cdot
\hX(\bz)}:\,\,.\labell{ver44}\eeqa  For graviton and dilaton the
polarization tensor $\veps_{ab}$ is symmetric, whereas,  for the
antisymmetric tensor this polarization is antisymmetric. All the
correlators above are simple to evaluate. The final result is
\beqa A(NS,NS,T,T)&\sim&\veps_{1ab}\veps_{2cd}\int
d^2z_1d^2z_2d^2z_3d^2z_4|z_{12}|^{-2}\prod_{i<j}^4|z_{ij}|^{2p_i\inn
p_j}\nonumber\\
&&\left[\frac{\eta^{ac}p_3\inn
p_4}{z_{12}z_{34}}-\frac{p_3^ap_4^c}{z_{13}z_{24}}+
\frac{p_4^ap_3^c}{z_{14}z_{23}}\right]\left[\frac{\eta^{bd}p_3\inn
p_4}{\bz_{12}\bz_{34}}-\frac{p_3^bp_4^d}{\bz_{13}\bz_{24}}+
\frac{p_4^bp_3^d}{\bz_{14}\bz_{23}}\right]\,\,.\nonumber\eeqa It
is easy to check that  the integrand has $SL(2,C)$ symmetry. One
should fix this symmetry by fixing position of three vertices at
 $z_1=\bz_1=\infty$, $z_2=\bz_2=0$, and $z_3=\bz_3=1$. After
this gauge fixing, one ends up with  one complex integral in the
$z$-plane. The imaginary part is zero and the real part is the
following (see the Appendix A): \beqa
A(NS,NS,T,T)&\sim&2\pi\left\{\left(p_3\inn\veps_1^T\inn\veps_2\inn
p_4+p_3\inn\veps_1\inn\veps_2^T\inn
p_4\right)\frac{\Ga(\frac{1}{2}-\frac{u}{2}
)\Ga(-\frac{s}{2})\Ga(\frac{1}{2}-\frac{t}{2})}{\Ga(\frac{3}{2}+\frac{u}{2})
\Ga(1+\frac{s}{2})\Ga(\frac{1}{2}+\frac{t}{2})}\right.\nonumber\\
&&\left.+p_3\inn\veps_1\inn p_3\,p_4\inn\veps_2\inn p_4
\frac{\Ga(-\frac{1}{2}-\frac{u}{2}
)\Ga(-\frac{s}{2})\Ga(\frac{1}{2}-\frac{t}{2})}{\Ga(\frac{3}{2}+\frac{u}{2})
\Ga(1+\frac{s}{2})\Ga(\frac{1}{2}+\frac{t}{2})}\right.\nonumber\\
&&\left.-p_3\inn\veps_1\inn p_4\,p_4\inn\veps_2\inn
p_3\frac{\Ga(\frac{1}{2}-\frac{u}{2}
)\Ga(-\frac{s}{2})\Ga(\frac{1}{2}-\frac{t}{2})}{\Ga(\frac{3}{2}+\frac{u}{2})
\Ga(1+\frac{s}{2})\Ga(\frac{3}{2}+\frac{t}{2})}\right.\nonumber\\
&&\left.-\frac{1}{2}\Tr(\veps_1^T\veps_2)
\frac{\Ga(\frac{1}{2}-\frac{u}{2}
)\Ga(-\frac{s}{2})\Ga(\frac{1}{2}-\frac{t}{2})}{\Ga(\frac{1}{2}+\frac{u}{2})
\Ga(1+\frac{s}{2})\Ga(\frac{1}{2}+\frac{t}{2})}+(4\leftrightarrow
3) \right\}\,\,,\labell{a14}\eeqa where the Mandelstam variables
are \beqa
s&=&-(p_3+p_4)^2\,\,,\nonumber\\
t&=&-(p_1+p_4)^2\,\,,\nonumber\\
u&=&-(p_2+p_4)^2\,\,.\labell{mandel}\eeqa They satisfy the
on-shell relation \beqa s+t+u=-2\,\,.\labell{m1}\eeqa  Note that
under $4\leftrightarrow 3$ the Mandelstam variables change as
$(s,t,u)\leftrightarrow (s,u,t)$. The term in the last line of
\reef{a14} has been also found in \cite{ikat}. If one of the NS
states is graviton and the other one is the antisymmetric two
tensor, then the whole amplitude vanishes. Moreover, when both NS
states are the antisymmetric two tensor, then the term in the
second line of \reef{a14} vanishes.   This indicates that there is
no linear tachyon coupling to two Kalb-Ramond states, because the
gamma functions in this term has a tachyonic pole.

Now to find the covariant expansion for the above amplitude, one
should note that  the covariant kinetic term produces massless
pole in $s$-channel and tachyonic pole in $t$-channel and
$u$-channel. Hence, one may send $s\rightarrow 0$ and
$t,u\rightarrow -1$. Fortunately, this limit is also consistent
with the constraint \reef{m1}. Hence, in this case the covariant
limit is simply \beqa \lim_{s\rightarrow 0\,, t,u\rightarrow
-1}A\labell{lim0}\eeqa One may use the constraint \reef{m1} to
rewrite the amplitude in such a way that the covariant limit
becomes $s,t,u\rightarrow 0$. To manage this, consider, for
example, the gamma functions in the first line of \reef{a14}. They
can be rewritten as the following: \beqa \lim_{s\rightarrow
0\,;t,u\rightarrow -1}\frac{\Ga(\frac{1}{2}-\frac{u}{2}
)\Ga(-\frac{s}{2})\Ga(\frac{1}{2}-\frac{t}{2})}{\Ga(\frac{3}{2}+\frac{u}{2})
\Ga(1+\frac{s}{2})\Ga(\frac{1}{2}+\frac{t}{2})}&=&\lim_{s,t,u\rightarrow
0}\frac{\Ga(1+\frac{s+t-u}{4}
)\Ga(-\frac{s}{2})\Ga(1+\frac{s+u-t}{4})}{\Ga(1+\frac{u-s-t}{4})
\Ga(1+\frac{s}{2})\Ga(\frac{t-u-s}{4})}\nonumber\eeqa Similar
thing can be done for all other gamma functions in \reef{a14}.
The final result is \beqa A
&\sim&2\pi\left\{\left(p_3\inn\veps_1^T\inn\veps_2\inn
p_4+p_3\inn\veps_1\inn\veps_2^T\inn
p_4\right)\frac{\Ga(1+\frac{s+t-u}{4}
)\Ga(-\frac{s}{2})\Ga(1+\frac{s+u-t}{4})}{\Ga(1+\frac{u-s-t}{4})
\Ga(1+\frac{s}{2})\Ga(\frac{t-u-s}{4})}\right.\nonumber\\
&&\left.+p_3\inn\veps_1\inn p_3\,p_4\inn\veps_2\inn p_4
\frac{\Ga(\frac{s+t-u}{4}
)\Ga(-\frac{s}{2})\Ga(1+\frac{s+u-t}{4})}{\Ga(1+\frac{u-s-t}{4})
\Ga(1+\frac{s}{2})\Ga(\frac{t-u-s}{4})}\right.\nonumber\\
&&\left.-p_3\inn\veps_1\inn p_4\,p_4\inn\veps_2\inn
p_3\frac{\Ga(1+\frac{s+t-u}{4}
)\Ga(-\frac{s}{2})\Ga(1+\frac{s+u-t}{4})}{\Ga(1+\frac{u-s-t}{4})
\Ga(1+\frac{s}{2})\Ga(1+\frac{t-u-s}{4})}\right.\nonumber\\
&&\left.-\frac{1}{2}\Tr(\veps_1^T\veps_2)
\frac{\Ga(1+\frac{s+t-u}{4}
)\Ga(-\frac{s}{2})\Ga(1+\frac{s+u-t}{4})}{\Ga(\frac{u-s-t}{4})
\Ga(1+\frac{s}{2})\Ga(\frac{t-u-s}{4})}+(4\leftrightarrow 3)
\right\}\,\,.\labell{a166}\eeqa In this form the covariant limit
\reef{lim0} becomes \beqa \lim_{s,t,u\rightarrow 0}A
\labell{lim00}\eeqa

The  S-matrix element of two NS and two massless scalar states
can also be written in the above form. To see this, consider the
latter amplitude which is given by the following correlation
function:\beqa A(NS,NS,g,g)&\sim& \langle:V_{(-1,-1)}^{\rm
NS}(p_1,\veps_1):V_{(-1,-1)}^{\rm NS}(p_2,\veps_2):V_{(0,0)}^{\rm
g}(p_3,\z_3):V_{(0,0)}^{\rm g}
(p_4,\z_4):\rangle\,\,,\nonumber\eeqa where the NS vertex operator
is given in \reef{ver44}, and the  scalar vertex operator is
given as \beqa V^g_{(0,0)}&\!\!\!\!=\!\!\!\!&\z_{ij}\int
d^2z:(\prt X^i(z)+ip\inn\psi(z)\psi^i(z))e^{ip\cdot
X(z)}:(\prt\hX^j(\bz)+ip\inn\hpsi(\bz)\hpsi^j(\bz))e^{ip\cdot
\hX(\bz)}:\,\,,\labell{ver444}\eeqa Straightforward calculation,
like what we have done before, gives the following final result:
\beqa
A(NS,NS,g,g)&\sim&2\pi\Tr(\z_3^T\z_4)\left\{\left(p_3\inn\veps_1^T\inn\veps_2\inn
p_4+p_3\inn\veps_1\inn\veps_2^T\inn
p_4\right)\frac{\Ga(1-\frac{u}{2}
)\Ga(-\frac{s}{2})\Ga(1-\frac{t}{2})}{\Ga(1+\frac{u}{2})
\Ga(1+\frac{s}{2})\Ga(\frac{t}{2})}\right.\nonumber\\
&&\left.+p_3\inn\veps_1\inn p_3\,p_4\inn\veps_2\inn p_4
\frac{\Ga(-\frac{u}{2}
)\Ga(-\frac{s}{2})\Ga(1-\frac{t}{2})}{\Ga(1+\frac{u}{2})
\Ga(1+\frac{s}{2})\Ga(\frac{t}{2})}\right.\nonumber\\
&&\left.-p_3\inn\veps_1\inn p_4\,p_4\inn\veps_2\inn
p_3\frac{\Ga(1-\frac{u}{2}
)\Ga(-\frac{s}{2})\Ga(1-\frac{t}{2})}{\Ga(1+\frac{u}{2})
\Ga(1+\frac{s}{2})\Ga(1+\frac{t}{2})}\right.\nonumber\\
&&\left.-\frac{1}{2}\Tr(\veps_1^T\veps_2) \frac{\Ga(1-\frac{u}{2}
)\Ga(-\frac{s}{2})\Ga(1-\frac{t}{2})}{\Ga(\frac{u}{2})
\Ga(1+\frac{s}{2})\Ga(\frac{t}{2})}+(4\leftrightarrow 3)
\right\}\,\,,\labell{a16666}\eeqa where the Mandelstam variables
satisfy \beqa s+t+u=0\,\,.\labell{m2}\eeqa Using this constraint,
it is easy to check explicitly that the above amplitude can be
rewritten in the  form appears in \reef{a166}. The only
difference is the extra factor of the scalar polarizations which
is one for the tachyon amplitude.

The covariant expansion of the amplitude \reef{a166} is \beqa
A&\!\!\!\!\sim\!\!\!\!&2\pi\left\{\left(p_3\inn\veps_1^T\inn\veps_2\inn
p_4+p_3\inn\veps_1\inn\veps_2^T\inn
p_4\right)\left(\frac{s+u-t}{2s}+\frac{\z(3)}{32}(s+u-t)(s^2-(t-u)^2)+\cdots\right)
\right.\nonumber\\
&&\left.+p_3\inn\veps_1\inn p_3\,p_4\inn\veps_2\inn p_4\left(
-\frac{2}{s}+\frac{4}{s+t-u}+\frac{\z(3)}{8}(s+u-t)^2+\cdots\right)
\right.\nonumber\\
&&\left.-p_3\inn\veps_1\inn p_4\,p_4\inn\veps_2\inn
p_3\left(-\frac{2}{s}-\frac{\z(3)}{8}(s^2-(t-u)^2)+\cdots\right)
\right.\nonumber\\
&&\left.-\frac{1}{2}\Tr(\veps_1^T\veps_2)\left(
-\frac{s^2-(t-u)^2}{8s}-\frac{\z(3)}{128}(s^2-(t-u)^2)^2+\cdots\right)
+(4\leftrightarrow 3) \right\}\,\,.\labell{a16}\eeqa  We shall
show that the first order terms above which has two momenta, are
reproduced in field theory by the action in which the covariant
tachyon kinetic term and the tachyon mass term are  added to the
standard low energy gravity action. The next leading terms have
eight momenta in which we are not interested in their field
theory couplings.

The S-matrix element of two dilatons and two tachyons can be read
from the general amplitude \reef{a166} by replacing the dilaton
polarization tensor in the amplitude. The dilaton polarization
tensor is $\veps^{ab}=(\eta^{ab}-p^a \ell^b-p^b
\ell^a)/\sqrt{D-2}$ where $p\inn \ell=1$. In the amplitude the
vector $\ell^a$ has to be canceled. This is a nontrivial check on
the amplitude \reef{a166}. Replacing this polarization tensor in
\reef{a166}, one finds, after some algebra,
 \beqa
A(\Phi',\Phi',T,T)&\sim&\frac{2\pi}{D-2}\left\{2p_3\inn
p_4\frac{\Ga(1+\frac{s+t-u}{4}
)\Ga(-\frac{s}{2})\Ga(1+\frac{s+u-t}{4})}{\Ga(1+\frac{u-s-t}{4})
\Ga(1+\frac{s}{2})\Ga(\frac{t-u-s}{4})}\right.\nonumber\\
&&\left.+p_3\inn p_3\,p_4\inn p_4 \frac{\Ga(\frac{s+t-u}{4}
)\Ga(-\frac{s}{2})\Ga(1+\frac{s+u-t}{4})}{\Ga(1+\frac{u-s-t}{4})
\Ga(1+\frac{s}{2})\Ga(\frac{t-u-s}{4})}\right.\nonumber\\
&&\left.-p_3\inn p_4\,p_4\inn p_3\frac{\Ga(1+\frac{s+t-u}{4}
)\Ga(-\frac{s}{2})\Ga(1+\frac{s+u-t}{4})}{\Ga(1+\frac{u-s-t}{4})
\Ga(1+\frac{s}{2})\Ga(1+\frac{t-u-s}{4})}\right.\nonumber\\
&&\left.-\left(\frac{D-2}{2}\right) \frac{\Ga(1+\frac{s+t-u}{4}
)\Ga(-\frac{s}{2})\Ga(1+\frac{s+u-t}{4})}{\Ga(\frac{u-s-t}{4})
\Ga(1+\frac{s}{2})\Ga(\frac{t-u-s}{4})}+(4\leftrightarrow 3)
\right\}\,\,.\labell{a1666}\eeqa Note that as expected the
auxiliary vector $\ell^a$ does not appear in the amplitude.
Expansion at $s,t,u\rightarrow 0$ gives, after some algebra, \beqa
A(\Phi',\Phi',T,T)&\sim&\frac{2\pi}{D-2}\left\{p_3\inn p_3
+p_3\inn p_3\,p_4\inn p_4\left( \frac{4}{s+t-u}\right)\right.\nonumber\\
&&\left.-\frac{D-2}{2}\left( -\frac{s^2-(t-u)^2}{8s}\right)+\cdots
+(4\leftrightarrow 3) \right\}\labell{a161}\,\,.\eeqa where dots
represent terms that start from $\z(3)$ order terms.

\subsection{Field theory analysis}
Now in field theory, we start with adding the covariant tachyon
kinetic term and an arbitrary tachyon mass term to the standard
low-energy gravity action  in D-dimensional space, \beqa
S_1^T&=&-\int
d^{D}x\sqrt{G}\left[e^{-2\Phi}(-2R-8\prt^a\Phi\prt_a\Phi+
\frac{3}{2}H_{abc}H^{abc}+
\frac{1}{2}\prt^aT\prt_aT+\frac{1}{2}m^2T^2\right]\,\,,\labell{action0}\eeqa
where  $H_{abc}=(\prt_aB_{bc}+\prt_c B_{ab}+\prt_b B_{ca})/3$. In
the Einstein frame ($G_{ab}=e^{\frac{4}{D-2}\Phi}g_{ab}$) it
becomes
 \beqa S_1^T&=&-\int
d^{D}x\sqrt{g}\left[-2R+\frac{1}{2}\prt^a\Phi'\prt_a\Phi'+
e^{\frac{-2\Phi'}{\sqrt{D-2}}}\left(\frac{3}{2}H_{abc}H^{abc}\right)
\right.\nonumber\\
&&\left.\qquad\qquad\qquad\qquad\qquad\qquad+
\frac{1}{2}\prt^aT\prt_aT+e^{\frac{1}{\sqrt{D-2}}\Phi'}
\left(\frac{1}{2}m^2T^2\right)\right]\labell{action}\eeqa where
dilaton $\Phi'=4\Phi/\sqrt{D-2}$, and graviton $h_{ab}$ is
related to the Einstein metric as $g_{ab}=\eta_{ab}+h_{ab}$. In
above field theory, we evaluate the  S-matrix element of two NS
fields and two tachyons.  Using the fact that particle $1,\,2$
are massless NS fields, and $3,\,4$ are tachyon with arbitrary
mass $m$, the Mandelstam variables \reef{mandel} become:  \beqa
s&=&-2p_1\inn p_2\,\,,\nonumber\\
t&=&-(-m^2+2p_2\inn p_3)\,\,,\labell{fmandel}\\
u&=&-(-m^2+2p_1\inn p_3)\,\,. \nonumber \eeqa Conservation of
momentum constrains them in the relation \beqa
s+t+u&=&2m^2\,\,.\labell{m11}\eeqa Note that if one restricts the
tachyon mass to the on-shell value $m^2=-1$, then above relation
reduces to the on-shell relation \reef{m1}.

Unlike in the string theory side that the S-matrix element for
graviton, Kalb-Ramond tensor, and dilaton   are given by a unique
amplitude \reef{a166},  in field theory side, one has to evaluate
each separately.  The $u$-channel amplitude for two tachyons and
two gravitons is given by the following Feynman rule: \beqa
A_u'(h,h,T,T)&=&\hV_{h_1T_3T}\hG_T\hV_{TT_4h_2}\,\,,\labell{a18}\eeqa
where the propagator and vertex function can be read from
\reef{action}. They are\beqa
\hG_T&=&\frac{i}{u-m^2}\,\,=\,\,\frac{-2i}{s+t-u}\,\,,\nonumber\\
\hV_{h_1T_3T}&=&i p_3\inn\veps_2\inn p_3\,\,.\nonumber\eeqa where
in the first line we have used the relation \reef{m11}.  Replacing
them in \reef{a18}, one finds \beqa
A_u'(h,h,T,T)&=&ip_3\inn\veps_1\inn p_3p_4\inn\veps_2\inn
p_4\left(\frac{2}{s+t-u}\right)\,\,.\nonumber\eeqa Note that the
$m$-dependence cancels out.  Comparing this with the corresponding
pole in the string theory amplitude \reef{a16}, one fins exact
agreement if the amplitude \reef{a166} is normalized by the factor
$i/(4\pi)$. The $t$-channel amplitude is the same as $u$-channel
in which $3\leftrightarrow 4$, which is obviously in agreement
with string theory.

The $s$-channel amplitude is given be the following Feynman rule:
\beqa A_s'(h,h,T,T)&=&(\hV_{h_1h_2h})^{ab}(\hG_h)_{ab}{}^{cd}
(\hV_{hT_3T_4})_{cd}\,\,,\labell{a17}\eeqa where the propagator,
 $\hV_{hT_3T_4}$, and $\hV_{h_1h_2h}$
can be read from the action \reef{action} (see \eg \cite{mg6}),
\beqa (\hG_h)^{ab,cd}&=&\frac{i}{2s}\left(\eta^{ac}\eta^{bd}+
\eta^{ad}\eta^{bc}-\frac{2}{D-2}\eta^{ab}\eta^{cd}\right)\,\,,\nonumber\\
(\hV_{hT_3T_4})^{ab}&=&-\frac{i}{2}\left[\eta^{ab}(m^2-p_3\inn
p_4)+p_3^a p_4^b+p_4^a p_3^b\right]\,\,,\labell{ver2}\\
(\hV_{h_1h_2h})^{ab}&\!\!=\!\!&-i\left[\left(\frac{3}{2}p_1\inn
p_2\eta^{ab}+p_1^{(a}p_2^{b)}-k^ak^b\right)\Tr(\veps_1\veps_2)-p_1\inn
\veps_2\inn\veps_1\inn
p_2\eta^{ab}+2p_2^{(a}\veps_2^{b)}\inn\veps_1\inn p_2
\right.\nonumber\\
&&\left.+2p_1^{(a}\veps_1^{b)}\inn\veps_2\inn p_1+
2p_1\inn\veps_2^{(a}\veps_1^{b)}\inn p_2-2p_1\inn
p_2\veps_1^{(a}\inn\veps_2^{b)}-p_1\inn\veps_2\inn
p_1\veps_1^{ab}-p_2\inn\veps_1\inn
p_2\veps_2^{ab}\right]\,\,,\nonumber\eeqa where $k=-(p_1+p_2)$ and
$p_1^{(a}p_2^{b)}$ means $(p_1^ap_2^b+p_1^bp_2^a)/2$. Replacing
them in \reef{a17}, after some algebra, one finds \beqa
A_s'(h,h,T,T)&\!\!\!=\!\!\!&i\left[p_3\inn\veps_1\inn\veps_2\inn
p_4\left(\frac{s+u-t}{2s}\right)-p_3\inn\veps_1\inn
p_3p_4\inn\veps_2\inn
p_4\left(\frac{1}{s}\right)\right.\nonumber\\
&&\left.+ p_3\inn\veps_1\inn p_4p_3\inn\veps_2\inn
p_4\left(\frac{1}{s}\right)-\Tr(\veps_1\veps_2)\left(\frac{(u-t)^2-s^2}{32s}\right)
\right.\labell{a19}\\
&&\left.-\frac{1}{8}\Tr(\veps_1\veps_2)s-\frac{1}{4}p_1\inn\veps_2\inn\veps_1\inn
p_2-\frac{1}{2}p_4\inn\veps_1\inn\veps_2\inn
p_4-\frac{1}{2}p_3\inn\veps_1\inn\veps_2\inn p_4+(3\leftrightarrow
4)\right]\,\,.\nonumber\eeqa Note that the D-dependence and
$m$-dependence cancel out. The massless poles are all in full
agreement with the string theory amplitude \reef{a16}. The left
over contact terms should be canceled by the $hhTT$ couplings of
field theory. Now the $hhTT$ contact terms in \reef{action} has
the following terms in momentum space: \beqa
\frac{i}{8}\Tr(\veps_1\veps_2)s+ip_3\inn\veps_1\inn\veps_2\inn
p_4+(3\leftrightarrow 4)\,\,.\nonumber\eeqa The two gravitons in
the first term above results from expanding the square root of
determinant of metric, and in the second term from expanding the
inverse of metric appearing in the kinetic term of the tachyon.
The above contact terms exactly cancel the contact terms in the
last line of \reef{a19}.

The $u$-channel and $t$-channel amplitude for scattering of two
tachyons and two Kalb-Ramond fields are zero, because there is no
vertex function with two tachyons and one Kalb-Ramond field in the
action \reef{action}. These vanishing amplitudes are consistent
with the string theory amplitude \reef{a16}, \eg $u$-channel
appears in the second line of \reef{a16} which is zero when both
NS states are Kalb-Ramond states. The $s$-channel amplitude is
given by the following Feynman rule: \beqa
A_s'(B,B,T,T)&=&(\hV_{B_1B_2h})^{ab}(\hG_h)_{ab}{}^{cd}
(\hV_{hT_3T_4})_{cd}+\hV_{B_1B_2\Phi'}\hG_{\Phi'}\hV_{\Phi'T_3T_4}\,\,,
\labell{a177}\eeqa where the propagators and vertex functions are
given in \reef{ver2} and in the following: \beqa
\hG_{\Phi'}&=&\frac{i}{s}\,\,,\nonumber\\
\hV_{\Phi'T_3T_4}&=&-\frac{im^2}{\sqrt{D-2}}\,\,,\labell{a1777}\\
\hV_{B_1B_2\Phi'}&=&
\frac{-2i}{\sqrt{D-2}}\left(\frac{}{}2p_1\inn\veps_2\inn\veps_1\inn
p_2-p_1\inn p_2\Tr(\veps_1\veps_2)\right)\,\,,\nonumber\\
(\hV_{B_1B_2h})^{ab}&=&-i\left[\frac{1}{2}\left(p_1\inn
p_2\eta^{ab}-2p_1^{(a}p_2^{b)}\right)\Tr(\veps_1\veps_2)-p_1\inn
\veps_2\inn\veps_1\inn
p_2\eta^{ab}+2p_1^{(a}\veps_2^{b)}\inn\veps_1\inn p_2
\right.\nonumber\\
&&\left.+2p_2^{(a}\veps_1^{b)}\inn\veps_2\inn p_1+
2p_1\inn\veps_2^{(a}\veps_1^{b)}\inn p_2-2p_1\inn
p_2\veps_1^{(a}\inn\veps_2^{b)}\right]\,\,.\nonumber\eeqa
Replacing them in \reef{a177}, after some algebra, one finds \beqa
A_s'(B,B,T,T)&=&-i\left[p_3\inn\veps_1\inn\veps_2\inn
p_4\left(\frac{s+u-t}{2s}\right)\right.\nonumber\\
&&\left.+ p_3\inn\veps_1\inn p_4p_3\inn\veps_2\inn
p_4\left(\frac{1}{s}\right)-\Tr(\veps_1\veps_2)\left(\frac{(u-t)^2-s^2}{32s}\right)
+(3\leftrightarrow 4)\right]\,\,,\nonumber\eeqa Note that again
all D-dependence and $m$-dependence cancel out. The above field
theory result  is in perfect agreement with the string result
\reef{a16}. Since there is no contact terms left over, one
concludes that the tachyon action has no coupling $H^2T^2$, as we
didn't include this in the action \reef{action0}. The absence of
the $H^2T^2$ coupling has been mentioned also in \cite{aatt2}.

The $u$-channel amplitude for scattering of two dilatons and two
tachyons in field theory \reef{action} is given by the following
Feynman rule: \beqa A_u'&=&\hV_{\Phi'_1T_3
T}\hG_T\hV_{TT_4\Phi'_2}\,\,\nonumber\\
&=&\left(\frac{2m^4}{D-2}\right)\frac{i}{s+t-u}\,\,,\nonumber\eeqa
which is in exact agreement with the string theory result in
\reef{a161} including the numerical factor. Note that the
amplitude \reef{a1666} has been already normalized by studying the
$u$-channel of graviton amplitude. The $t$-channel amplitude
which can be read from the $u$-channel amplitude above by
interchanging $3\leftrightarrow 4$ is agree with string theory
result for obvious reason. The $s$-channel is given by the
following amplitude: \beqa A_s'&=&
(\hV_{\Phi'_1\Phi'_2h})^{ab}(\hG_h)_{ab}{}^{cd}(\hV_{hT_3T_4})^{cd}\,\,,
\labell{a77}\eeqa where the propagator and vertex functions are
given in \reef{ver2} and  \beqa
(\hV_{h\Phi'_1\Phi'_2})^{ab}&=&-\frac{i}{2}\left[\eta^{ab}\left(-p_1\inn
p_2\right)+p_1^ap_2^b+p_1^bp_2^a\right]\,\,.\nonumber\eeqa
Replacing them in \reef{a77}, one finds \beqa
A_s'&=&-\frac{i}{4s}\left(-2p_1\inn p_2 p_3\inn p_4+2m^2p_1\inn
p_2+2p_1\inn p_3 p_2\inn
p_3+2p_1\inn p_4 p_2\inn p_3\right)\nonumber\\
&=&\frac{i}{16s}\left(s^2-(t-u)^2\right)\,\,.\nonumber\eeqa
Comparing this with the massless pole in \reef{a161}, one finds
exact agreement. Finally, the contact term in \reef{a161} is
exactly the coupling $TT\Phi'\Phi'$ in the action \reef{action}.

The leading terms of the amplitude \reef{a166} at the covariant
limit \reef{lim00},  is then consistent with the covariant action
\reef{action0}. As we already mentioned in the previous section,
the  amplitude \reef{a166} is also describe the S-matrix element
of two massless scalars $g$ and two NS states. Accordingly, the
action \reef{action0} is also action for the scalar field as
well. For the scalar case the mass term is of course zero, \ie,
 \beqa S_1^g&=&-\int
d^{D}x\sqrt{G}\left[e^{-2\Phi}(-2R-8\prt^a\Phi\prt_a\Phi+
\frac{3}{2}H_{abc}H^{abc}+ \frac{1}{2}\prt^a g\prt_a g
\right]\,\,,\labell{action00}\eeqa

\section{Four tachyon amplitude}
We repeat the same analysis as in previous section for four
tachyons. So we begin with the evaluation of the  sphere 4-point
function of four tachyon vertex operators. This amplitude is
given by the following correlation function: \beqa
A(T,T,T,T)&\sim& \langle:V_{(0,0)}^{\rm T}(p_1):V_{(0,0)}^{\rm
T}(p_2):V_{(-1,-1)}^{\rm T}(p_3):V_{(-1,-1)}^{\rm T}
(p_4):\rangle\,\,,\nonumber\eeqa where we have used the tachyon
vertex operators in different pictures  in order to saturate the
background supercharge of the sphere.  The vertex operator in
(0,0) picture is given in \reef{ver44}, and in (-1,-1) picture is
given in the following : \beqa V_{(-1,-1)}^{\rm T}(p)&=&\int
d^z:e^{-\phi(z)}e^{ip\cdot X(z)}:e^{-\hp(\bz)}e^{ip\cdot
\hX(\bz)}\,\,.\nonumber \eeqa These correlators have been
evaluated in \cite{ikat},  \beqa A(T,T,T,T)&\sim&(p_1\inn
p_2)^2\int d^2z|z|^{2p_2\cdot p_4}|1-z|^{2p_3\cdot
p_4-2}\,\,,\nonumber\\
&\sim&2\pi\frac{\Ga(-\frac{u}{2})\Ga(-\frac{s}{2})\Ga(-\frac{t}{2})}
{\Ga(1+\frac{u}{2})\Ga(1+\frac{s}{2})\Ga(1+\frac{t}{2})}\,\,,\labell{a11}
\eeqa where the Mandelstam variables are those appearing in
\reef{mandel}. In this case they satisfy the on-shell
relation\beqa s+t+u=-4\,\,.\labell{mandel2}\eeqa

To find the covariant expansion of this amplitude, one should
realize that the tachyon covariant kinetic term produce massless
poles in all channels. However, the constraint \reef{mandel2}
does not allow us to sent all $s,t,u$ to zero at the same time,
\ie $s,t,u\rightarrow 0$, to produce the massless poles. Following
\cite{mrga}, to find the correct way of expanding the tachyon
amplitude, one should compare the tachyon amplitude with the
S-matrix element of four massless scalars. In that case it is
known how to expand the amplitude to produce the Feynman
amplitude resulting from scalar kinetic term. Following the same
steps, one can find the covariant expansion of the tachyon
amplitude \reef{a11}.

The S-matrix element of four scalars is given by the following
correlation function: \beqa A(g,g,g,g)&\sim&
\langle:V_{(0,0)}^{\rm g}(p_1,\z_1):V_{(0,0)}^{\rm
g}(p_2,\z_2):V_{(-1,-1)}^{\rm g}(p_3,\z_3):V_{(-1,-1)}^{\rm g}
(p_4,\z_4):\rangle\,\,,\nonumber\eeqa where the scalar vertex
operator in (0,0) picture is given in \reef{ver444}, and in
(-1,-1) picture is given by
 \beqa
V^g_{(-1,-1)}&\!\!\!\!=\!\!\!\!&\z_{ij}\int
d^2z:e^{-\phi(z)}\psi^i(z)e^{ip\cdot
X(z)}:e^{-\hp(\bz)}\hpsi^j(\bz)e^{ip\cdot
\hX(\bz)}:\,\,.\labell{ver4}\nonumber\eeqa Straightforward
evaluation of the correlators gives the result \beqa
A(g,g,g,g)&\sim&\z_{1ij}\z_{2kl}\z_{3nm}\z_{4pq}\int d^2z_1
d^2z_2d^2z_3d^2z_4 \prod_{a<b}^4 |z_{ab}|^{2p_a\cdot
p_b}\nonumber\\
&&\times\left[\frac{\eta^{ik}\eta^{np}(1-p_1\inn
p_2)}{z_{12}^2z_{34}^2}+\frac{\eta^{in}\eta^{kp}p_1\inn
p_2}{z_{12}z_{13}z_{24}z_{34}}-\frac{\eta^{ip}\eta^{kn}p_1\inn
p_2}{z_{12}z_{14}z_{23}z_{34}}\right]\nonumber\\
&&\times\left[\frac{\eta^{jl}\eta^{mq}(1-p_1\inn
p_2)}{\bz_{12}^2\bz_{34}^2}+\frac{\eta^{jm}\eta^{lq}p_1\inn
p_2}{\bz_{12}\bz_{13}\bz_{24}\bz_{34}}-\frac{\eta^{jq}\eta^{lm}p_1\inn
p_2}{\bz_{12}\bz_{14}\bz_{23}\bz_{34}}\right]\,\,.\nonumber\eeqa
It is easy to check that  the integrand has $SL(2,C)$ symmetry.
One should fix this symmetry by fixing position of three vertices
at
 $z_1=\bz_1=\infty$, $z_2=\bz_2=0$, and $z_3=\bz_3=1$. After
this gauge fixing, one ends up with  one complex integral in the
$z$-plane. The imaginary part is zero and the real part is the
following (see the Appendix A): \beqa
A(g,g,g,g)&\sim&2\pi\left(\Tr(\z_1^T\z_2)\Tr(\z_3^T\z_4)
\frac{\Ga(1-\frac{u}{2})\Ga(-\frac{s}{2})\Ga(1-\frac{t}{2})}
{\Ga(\frac{u}{2})\Ga(1+\frac{s}{2})\Ga(\frac{t}{2})}\right.\nonumber\\
&&\left.-\Tr(\z_1^T\z_2\z_3^T\z_4)
\frac{\Ga(1-\frac{u}{2})\Ga(1-\frac{s}{2})\Ga(1-\frac{t}{2})}
{\Ga(\frac{u}{2})\Ga(1+\frac{s}{2})\Ga(1+\frac{t}{2})}\right.\nonumber\\
&&\left.-\Tr(\z_1^T\z_2\z_4^T\z_3)
\frac{\Ga(1-\frac{u}{2})\Ga(1-\frac{s}{2})\Ga(1-\frac{t}{2})}
{\Ga(1+\frac{u}{2})\Ga(1+\frac{s}{2})\Ga(\frac{t}{2})}\right.\nonumber\\
&&\left. +(1234\rightarrow 1324)+(1234\rightarrow
1432)\frac{}{}\right)\,\,,\labell{a10}\eeqa where the Mandelstam
variables satisfy \reef{m2}. Note that under $(1234)\rightarrow
(1324)$ the Mandelstam variables change as $(s,t,u)\rightarrow
(u,t,s)$, and under $(1234)\rightarrow (1432)$ change as
$(s,t,u)\rightarrow (t,s,u)$.

Now, the gamma functions in the second and third line of
\reef{a10} has no massless pole at all. This indicates that these
terms have no contribution in producing the massless poles. In
other words, only the first line produces the massless pole in
$s$-channel resulting from covariant kinetic term. In this case,
the expansion is of course at $s,t,u\rightarrow 0$. So at the
covariant limit the gamma function should be send to \beqa
\frac{\Ga(1-\frac{u}{2})\Ga(-\frac{s}{2})\Ga(1-\frac{t}{2})}
{\Ga(\frac{u}{2})\Ga(1+\frac{s}{2})\Ga(\frac{t}{2})}&\rightarrow&
\frac{\Ga(1)\Ga(0)\Ga(1)} {\Ga(0)\Ga(1)\Ga(0)}\nonumber\eeqa
Similarly for $t$ and $u$ channels. Using these results from the
scalar amplitude, one realizes immediately that the tachyon
amplitude \reef{a11} should be rewritten as
$A(T,T,T,T)=(A+A+A)/3$, and the covariant limit is the following:
\beqa s-{\rm
channel}:&&\lim_{s\rightarrow 0\,,t,u\rightarrow -2}A\nonumber\\
t-{\rm channel}:&&\lim_{t\rightarrow 0\,,s,u\rightarrow
-2}A\nonumber\\
u-{\rm channel}:&&\lim_{u\rightarrow 0\,,s,t\rightarrow
-2}A\labell{lim1}\eeqa It may seems strange that in this limit
one should send $s$, say, once to zero and once to -2. However,
this happens only in the particular form of the amplitude
\reef{a11}.  Using the constraint \reef{mandel2}, one can rewrite
the amplitude in the following  form:  \beqa A
&\sim&\frac{2\pi}{3}\left(
\frac{\Ga(1+\frac{s+t-u}{4})\Ga(-\frac{s}{2})\Ga(1+\frac{s+u-t}{4})}
{\Ga(\frac{u-s-t}{4})\Ga(1+\frac{s}{2})\Ga(\frac{t-u-s}{4})}\right.\nonumber\\
&&\left.\qquad\qquad\qquad+(1234\rightarrow 1324)+(1234\rightarrow
1432)\frac{}{}\right)\labell{a12}\eeqa In this  form, the
covariant limit is $s,t,u\rightarrow 0$.  It is easy to check
that using the constraint \reef{m2}, the terms in the first line
of \reef{a10} can also be rewritten in the  form appearing in
\reef{a12}. The scalar amplitude \reef{a10} has some other terms
that the tachyon amplitude does not have them. This indicates
that four scalars couplings and the four tachyon couplings that
their coefficients are independent of the tachyon mass, are not
the same. The scalar couplings has the tachyon couplings  as well
as some other couplings that result from expansion of the terms
in the second and third line of \reef{a10}. Expansion of
\reef{a12} at $s,t,u\rightarrow 0$ gives the following leading
terms: \beqa A(T,T,T,T)&\sim&\frac{2\pi}{3}\left(\left(
\frac{(t-u)^2-s^2}{8s}
-\frac{\z(3)}{28}(s^2-(t-u)^2)^2+\cdots\right)\right.\nonumber\\
&&\left. +(1234\rightarrow 1324)+(1234\rightarrow
1432)\frac{}{}\right)\,\,,\labell{a13}\eeqa The first order terms
above should be reproduced by  the two derivative action
\reef{action}.
\subsection{Field theory analysis}

Now in field theory,  using the fact that particles  are all
tachyon with arbitrary mass $m$, the Mandelstam variables
\reef{mandel} become: \beqa
s&=&-(-2m^2+2p_1\inn p_2)\,\,,\nonumber\\
t&=&-(-2m^2+2p_2\inn p_3)\,\,,\nonumber\\
u&=&-(-2m^2+2p_1\inn p_3)\,\,. \labell{fmandel2}\nonumber \eeqa
Conservation of momentum constrains them in the relation \beqa
s+t+u&=&4m^2\,\,.\labell{fconstraint3}\nonumber\eeqa Here again if
one restricts the mass of tachyon to on-shell value $m^2=-1$, the
above relation also reduces to the on-shell relation
\reef{mandel2}.

The $s$-channel amplitude in field theory \reef{action} is given
by the following Feynman rule: \beqa A_s'&=&(\hV_{T_1T_2
h})^{ab}(\hG_h)_{ab}{}^{cd}(\hV_{hT_3T_4})_{cd}+ \hV_{T_1 T_2
\Phi'}\hG_{\Phi'}\hV_{\Phi'T_3T_4}\,\,,\labell{a66}\nonumber\eeqa
where the vertex functions and propagators appear in \reef{ver2}
and \reef{a1777}.  Replacing them in above equation, one finds
\beqa A_s'&=&-\frac{i}{4s}\left(-2(m^2-p_1\inn p_2)(m^2-p_3\inn
p_4)+2p_1\inn p_3 p_2\inn p_4+2p_1\inn p_4 p_2\inn
p_3\right)\nonumber\\
&=&-\frac{i}{16s}\left((u-t)^2-s^2\right)\,\,.\nonumber\eeqa Note
that all $D$-dependence and $m$-dependence cancels out. Now
comparing this field theory massless pole with the massless pole
of string theory amplitude \reef{a13}, one finds exact agreement
if normalizes the string amplitude \reef{a12} by factor
$-3i/(4\pi)$. The $t$-channel and $u$-channel calculation in field
theory can be read from the $s$-channel amplitude by replacing
$(1234)\rightarrow (1432)$ and $(1234)\rightarrow (1324)$,
respectively. They obviously agree with string theory amplitude
\reef{a13}. Hence, there is no contact term left over at this
order. This indicates that there is no $m^4T^4$ coupling in the
type 0 theory\footnote{One may object that the next leading terms
of \reef{a13} may have $T^4$ coupling. This is very unlikely
because the coefficient of this term is $\z(3)$, and the
$\alpha'$ order of this term tell us that even if this term
produce $T^4$ coupling, its coefficient would be $m^8$.}. This is
unlike the open string case \cite{mrga}, that the tachyon
potential has $m^4T^4$ coupling. Therefore the tachyon potential
in type 0 theory is \beqa
V(T)&=&\frac{1}{2}m^2T^2\,\,.\labell{a21}\eeqa When the tachyon
mass  is on-shell, the potential has a maximum at $T=0$
($V(0)=0$) and unbounded minimum at $T=\pm\infty$
($V(\pm\infty)=-\infty$). In the sigma model approach
\cite{aatt}, on the other hand,  a bounded potential has been
found for tachyon \cite{oa}. However, one expects that the two
approach are not using the same field variables, \ie, the two
results should be related to each other by some field
redefinition.

The  amplitude \reef{a12} describes also one part of the S-matrix
element of four scalar \reef{a10}, \ie, the terms in the first
line of \reef{a10}. The terms in the second and third lines of
\reef{a10} when the polarization of the scalars are replaced by
1,  have zero contribution to the leading two derivative terms.
Hence, the leading term in \reef{a13} describes  the leading
terms of the four scalar amplitude as well. Accordingly, the
action \reef{action00} is consistent with the leading terms of
four scalar amplitude \reef{a10}.

\section{Two tachyon-two RR  amplitude}\label{scatt}

We start  this section by  evaluating the  sphere  4-point
function of two RR and two tachyon vertex operators. This
amplitude for the case of RR scalar has been found in
\cite{ikat}. The amplitude for arbitrary RR state may be  given by
the following correlation function: \beqa
A(C,C,T,T)&\!\!\!\sim\!\!\!& \langle:V_{(-1/2,-1/2)}^{\rm
RR}(p_1,\veps_1):V_{(-1/2,-1/2)}^{\rm
RR}(p_2,\veps_2):V_{(-1,0)}^{\rm T}(p_3):V_{(0,-1)}^{\rm T}
(p_4):\rangle\,\,,\labell{amp}\eeqa where $\veps$'s are
polarization of the RR fields and $p$'s are momentum of states.
The vertex operators  are : \beqa V^{\rm
RR}_{(-1/2,-1/2)}(p,\veps)&=&\int
d^2z\,(P_{\mp}\Ga_{(n)})^{AB}:e^{-\phi(z)/2}S_A(z)e^{ip\cdot X}:
e^{-\hp(\bz)/2}\hS_B(\bz)e^{ip\cdot \hX(\bz)}:\,\,,\nonumber\\
V_{(-1,0)}^{\rm T}(p)&=&\int d^2z:e^{-\phi(z)}e^{ip\cdot
X(z)}:ip\inn\hpsi(\bz)e^{ip\cdot \hX(\bz)}:\,\,,\nonumber\\
V_{(0,-1)}^{\rm T}(p)&=&\int d^2z:ip\inn\psi(z)e^{ip\cdot
X(z)}:e^{-\hp(\bz)}e^{ip\cdot \hX(\bz)}:\,\,,\labell{ver3}\eeqa
where $P_{\mp}$ are the two different chiral projection operators
that refer to two different set of RR states. The RR polarization
tensor $\veps_{a_1a_2\cdots a_{n-1}}$ is included in $\Ga_{(n)}$.
We refer the reader to ref.\cite{mg6} for this relation and for
our other conventions. The on-shell conditions for RR fields are
$p^2=0=\veps\inn p$ and for tachyon is $p^2=1$.

In evaluating the correlators in \reef{amp}, one needs the
correlation of two spin operators and one world-sheet fermion
that is given by (see, \eg \cite{jp}) \beqa
<:S_A(z_1):S_B(z_2):\psi^{\mu}(z_3):>&=&
\frac{1}{\sqrt{2}}(\ga^{\mu})_{AB}z_{12}^{-3/4}(z_{13}z_{23})^{-1/2}\,\,.
\labell{corr}\eeqa The other correlators in \reef{amp} can easily
be evaluated, using different world-sheet propagators \cite{jp}.
Performing these correlations, one finds that the integrand has
$SL(2,C)$ symmetry. One should fix this symmetry by fixing
position of three vertices at say $z_1=\bz_1=\infty$,
$z_2=\bz_2=0$, and $z_3=\bz_3=1$. After this gauge fixing, one
ends up with only one real integral in the $z$-plane, \beqa
A&\sim&\alpha\int d^2z|z|^{2p_2\cdot p_4-1}|1-z|^{2p_3\cdot
p_4}\,\,,\labell{a1}\eeqa where $\alpha$ includes the kinematic
factors, \beqa
\alpha&=&\frac{1}{2}(P_{\mp}\Ga_{1(n)})^{AB}(P_{\mp}\Ga_{2(n)})^{CD}
(ip_4\inn\ga)_{AC}(ip_3\inn\ga)_{BD}\labell{alpha}\\
&=&-\frac{8}{n!}\left(p_3\inn p_4F_1^{a_1\cdots a_n}
F_{2a_1\cdots a_n}-np_{3a} F_1^{a a_2\cdots a_n} F_{2b a_2\cdots
a_n} p_4^b-np_{4a}F_1^{a a_2\cdots a_n} F_{2b a_2\cdots a_n}
p_3^b\right)\,\,. \nonumber\eeqa We refer the reader to \cite{mg6}
for our convention for the gamma matrices.

The integral in amplitude \reef{a1} can be performed, and the
result in terms of the Mandelstam variables is (see the Appendix A
)\beqa A&\sim&2\pi
\alpha\frac{\Ga(-\frac{u}{2})\Ga(-\frac{s}{2})\Ga(-\frac{t}{2})}
{\Ga(1+\frac{u}{2})\Ga(1+\frac{s}{2})\Ga(1+\frac{t}{2})}\,\,,\labell{a2}\eeqa
where the Mandelstam variables satisfy the constrain \reef{m1}.
The above result for the case that $n=1$ has been found in
\cite{ikat}. Note that the amplitude has symmetry
$3\leftrightarrow 4$ and $1\leftrightarrow 2$.

The amplitude \reef{a2} has massless poles in all channels.
However, only the massless pole in the $s$-channel can be
reproduced by covariant kinetic term. The massless poles in other
channels are reproduced by assuming that the field theory has
$F\bF T$ coupling.  To find the covariant expansion of the
amplitude, one should send $s\rightarrow 0$ to produce the
massless pole resulting from covariant kinetic term, and send
$t\rightarrow 0$ ($u\rightarrow 0$) to produce the massless pole
resulting from the covariant coupling $F\bF T$ in the $t$-channel
($u$-channel). The constraint \reef{m1} does not allow  all
$s,t,u$ approach zero all at the same time.

Using the fact that the coefficient $\alpha$ in \reef{a2} has
four momenta, the expansion of the gamma function at the
covariant limit should have constant massless poles. Moreover,
the limit should keep the symmetry of the amplitude. Using these,
one may rewrite  the amplitude  as $A(C,C,T,T)=(A+A)/2$. Then,
the covariant limit is \beqa s,t-{\rm
channel}:&&\lim_{s,t\rightarrow 0\,,u\rightarrow -2}A\nonumber\\
s,u-{\rm channel}:&&\lim_{s,u\rightarrow 0\,,t\rightarrow
-2}A\labell{lim11}\eeqa Again using the constraint \reef{m1}, one
can rewrite the amplitude as \beqa A(C,C,T,T)&\sim&\pi\alpha
\left(\frac{\Ga(1+\frac{s+t}{2})\Ga(-\frac{s}{2})\Ga(-\frac{t}{2})}
{\Ga(-\frac{s+t}{2})\Ga(1+\frac{s}{2})\Ga(1+\frac{t}{2})}
\right.\nonumber\\
&&\left.\qquad\qquad
+\frac{\Ga(-\frac{u}{2})\Ga(-\frac{s}{2})\Ga(1+\frac{s+u}{2})}
{\Ga(1+\frac{u}{2})\Ga(1+\frac{s}{2})\Ga(-\frac{s+u}{2})}\right)\,\,.\labell{a44}
\eeqa  In this  form, the covariant limit is $s,t,u\rightarrow
0$. Expansion at this limit is \beqa A(C,C,T,T)
&\!\!\!\sim\!\!\!&\pi\alpha\left(\frac{}{}-\frac{4}{s}-\frac{2}{t}-\frac{2}{u}+
\frac{\z(3)}{2}((s+t)^2+(s+u)^2)+\cdots\right)\labell{expan1}\eeqa

One may object that there might be other limit than \reef{lim11}
for the amplitude that produces massless poles in all channels.
One may write  $A=(A+A+A)/3$ and send \beqa s,t-{\rm
channel}:&&\lim_{s,t\rightarrow 0\,,u\rightarrow -2}A\nonumber\\
s,u-{\rm channel}:&&\lim_{s,u\rightarrow 0\,,t\rightarrow
-2}A\labell{lim22}\\
t,u-{\rm channel}:&&\lim_{t,u\rightarrow 0\,, s\rightarrow -2}A
\nonumber\eeqa or send \beqa s-{\rm
channel}:&&\lim_{s\rightarrow 0\,,t,u\rightarrow -1}A\nonumber\\
u-{\rm channel}:&&\lim_{u\rightarrow 0\,,s,t\rightarrow
-1}A\labell{lim222}\\
t-{\rm channel}:&&\lim_{t\rightarrow 0\,, s,u\rightarrow -1}A
\nonumber\eeqa They all would produce massless poles in all $s$-,
$t$- and $u$-channels. However,  the coefficient of the massless
pole in the $s$-channel would  not be the same as \reef{expan1}.
On the other hand, we know the coefficient of $s$-channel is fixed
because it related to the standard covariant kinetic term. To
show that the coefficient of the $s$-channel in \reef{expan1} is
the correct one, accordingly the covariant limit \reef{lim11} is
the only correct  limit, we compare the result with the
$s$-channel of the scalar amplitude. The $s$-channel pole of the
scalar amplitude must be exactly the same as the tachyon
amplitude, as both are related to standard kinetic terms.

The S-matrix element of two RR and two massless scalar vertex
operators is given by correlation \reef{amp} in which the tachyon
vertex operators are replaced by the following  scalar vertex
operators: \beqa
V^g_{(-1,0)}(p_3,\z_3)&\!\!\!\!=\!\!\!\!&\z_{3ij}\int
d^2z:e^{-\phi(z)}\psi^i(z)e^{ip_3\cdot
X(z)}:(\prt\hX^j(\bz)+ip_3\inn\hpsi(\bz)\hpsi^j(\bz))e^{ip_3\cdot
\hX(\bz)}:\,\,,\nonumber\\
V^g_{(0,-1)}(p_4,\z_4)&\!\!\!\!=\!\!\!\!&\z_{4kl}\int d^2z:(\prt
X^k(z)+ip_4\inn\psi(z)\psi^k(z))e^{ip_4\cdot
X(z)}:e^{-\hp(\bz)}\hpsi^l(\bz)e^{ip_4\cdot
\hX(\bz)}:\,\,,\labell{vertex}\eeqa   The necessary correlation
functions between the world-sheet fermions and the spin operators
appearing in this amplitude  is \beqa
<:S_A(z_1):A_B(z_2):\psi^i(z_3):ip_4\inn\psi(\z_4)\psi^k(z_4):>\,\,,
\nonumber\eeqa which can be reduced to the correlation
\reef{corr} using the following relations (see \eg \cite{jp}):
\beqa :S_A(z_1):ip_4\inn\psi(z_4)\psi^k(z_4):&\sim&\frac{1}{4}
(ip_4\inn\ga\ga^k-\ga^k
ip_4\inn\ga)_A{}^{A'}S_{A'}(z_1)z_{14}^{-1}\,\,,\nonumber\\
:\psi^i(z_3):ip_4\inn\psi(z_4)\psi^k(z_4):&\sim&
\eta^{ik}ip_4\inn\psi(z_3)z_{34}^{-1}\,\,,\nonumber\eeqa where in
the second line we have used the fact that momentum is in
non-compact space and the indices $i$ is  in the orthogonal
compact space, \ie $\eta^{ai}=0$. This property simplifies
greatly the evaluation of the correlation functions in
$A(C,C,g,g)$. The final result is \beqa
A(C,C,g,g)&\sim&\frac{1}{8}\z_{3ij}\z_{4kl}(P_{\mp}
\Ga_{1(n)})^{AB}(P_{\mp}\Ga_{2(n)})^{CD} \int
d^2z_1d^2z_2d^2z_3d^2z_4\prod_{m<n}^4 |z_{nm}|^{2p_n\cdot
p_m}\nonumber\\
&&\times \left[(ip_4\inn\ga\ga^k\ga^i)_{AC}
(z_{23}z_{12}z_{34}z_{14})^{-1}+(ip_4\inn\ga\ga^k\ga^i)_{CA}
(z_{13}z_{12}z_{34}z_{24})^{-1}\right]\nonumber\\
&&\times\left[(ip_3\inn\ga\ga^l\ga^j)_{BD}(\bz_{24}\bz_{12}\bz_{43}\bz_{13})^{-1}+
(ip_3\inn\ga\ga^l\ga^j)_{DB}(\bz_{14}\bz_{12}\bz_{43}\bz_{23})^{-1}\right]\,\,.
\nonumber\eeqa The integrand is $SL(2,C)$ invariant. Fixing this
symmetry, like in the tachyon case, one finds a complex integral
in the $z$-plane. The imaginary part of the integral is zero and
the real part is the following (see the Appendix A): \beqa
A(C,C,g,g)&\sim&\frac{\pi\alpha}{2}
\left(\left(\Tr(\z_3^T\z_4)-\Tr(\z_3\z_4)+\Tr(\z_3)\Tr(\z_4)\right)
\frac{\Ga(1-\frac{u}{2})\Ga(-\frac{s}{2})\Ga(-\frac{t}{2})}
{\Ga(\frac{u}{2})\Ga(1+\frac{s}{2})\Ga(1+\frac{t}{2})}
\right.\labell{a3}\\
&&\left.+2\left(\Tr(\z_3^T\z_4)+\Tr(\z_3\z_4)-\Tr(\z_3)\Tr(\z_4)\right)
\frac{\Ga(1-\frac{u}{2})\Ga(-\frac{s}{2})\Ga(1-\frac{t}{2})}
{\Ga(1+\frac{u}{2})\Ga(1+\frac{s}{2})\Ga(1+\frac{t}{2})}\right.\nonumber\\
&&\left.
+\left(\Tr(\z_3^T\z_4)-\Tr(\z_3\z_4)+\Tr(\z_3)\Tr(\z_4)\right)
\frac{\Ga(-\frac{u}{2})\Ga(-\frac{s}{2})\Ga(1-\frac{t}{2})}
{\Ga(1+\frac{u}{2})\Ga(1+\frac{s}{2})\Ga(\frac{t}{2})}\right)\,\,.\nonumber
\eeqa   Note that the result has the expected symmetry between
3,4 and between 1,2.  The Mandelstam variables are given in
\reef{mandel}, and they satisfy  the relation \reef{m2}.

If one considers the case that there is only one scalar, \ie the
compact space is circle, then the polarization factors simplify
to 1. Then, expansion of this amplitude at low energy
$s,t,u\rightarrow 0$ gives the following leading terms: \beqa
A(C,C,g,g)&\sim&\frac{\pi\alpha}{2}\left(-\frac{8}{s}-\frac{2}{t}-
\frac{2}{u}+\cdots\right)\,\,,\labell{a5}\eeqa The coefficient of
massless pole in the $s$-channel is exactly the same as the
massless pole in the tachyon amplitude \reef{expan1}. This
confirms that the covariant limit \reef{lim11} is the only
correct covariant limit of the tachyon amplitude \reef{a2}, \ie
the limits \reef{lim22}, \reef{lim222} although consistent with
the constraint \reef{m1}, they are not correct covariant limits.

\subsection{Field theory analysis}

Now in field theory, consider adding the following couplings to
the action \reef{action}: \beqa S_2^T&=&-\int
d^{D}x\sqrt{G}\left[\frac{1}{2}(F_{(n)}\inn
F_{(n)}+\bF_{(n)}\inn\bF_{(n)})(1+b_TT^2)+F_{(n)}\inn\bF_{(n)}(a_T
T)\right]\,\,,\labell{action10}\eeqa where $F_{(n)}\inn
F_{(n)}\equiv\frac{1}{n!}F^{a_1\cdots a_n}F_{a_1\cdots a_n}$. The
above action is parametrized by two constants $a_T,b_T$. In the
Einstein frame  it becomes
 \beqa S_2^T&=&-\int
d^{D}x\sqrt{g}\left[e^{\frac{D-2n}{2\sqrt{D-2}}\Phi'}\left(\frac{1}{2}(F_{(n)}\inn
F_{(n)}+\bF_{(n)}\bF_{(n)})(1+b_TT^2)+ F_{(n)}\inn\bF_{(n)}(a_T
T)\right)\right]\,\,,\nonumber\eeqa In $S_1^T+S_2^T$ field theory,
we evaluate the  S-matrix element of two RR fields and two
tachyons.  Using the fact that particle $1,\,2$ are massless RR
fields, and $3,\,4$ are tachyon with arbitrary mass $m$, the
Mandelstam variables \reef{mandel} turn into \reef{fmandel}.

The $s$-channel amplitude is given by the following Feynman rule:
\beqa
A_s'&=&(\hV_{F_1F_2h})^{ab}(\hG_h)_{ab}{}^{cd}(\hV_{hT_3T_4})_{cd}+
\hV_{F_1F_2\Phi'}\hG_{\Phi'}\hV_{\Phi'T_3T_4}\,\,,\labell{a6}\eeqa
the vertex functions and propagators are given in \reef{ver2},
\reef{a1777}, and in the following:
 \beqa
\hV_{\Phi'F_1F_2}&=&-i\frac{D-2n}{2\sqrt{D-2}}F_{1(n)}\inn
F_{2(n)}\,\,,\nonumber\\
(\hV_{hF_1F_2})^{ab}&=&-\frac{i}{2n!}\left[\eta^{ab}F_{1a_1\cdots
a_n}F_2^{a_1\cdots a_n}-nF_1^a{}_{a_2\cdots a_n}F_2^{ba_2\cdots
a_n}-nF_1^b{}_{a_2\cdots a_n}F_2^{a a_2\cdots
a_n}\right]\,\,.\nonumber\eeqa Replacing them in \reef{a6}, one
finds, after some simple algebra, \beqa
A'_s&=&\frac{-i}{2n!s}\left(p_3\inn p_4F_1^{a_1\cdots a_n}
F_{2a_1\cdots a_n}-np_{3a} F_1^{a a_2\cdots a_n} F_{2b a_2\cdots
a_n} p_4^b-np_{4a}F_1^{a a_2\cdots a_n} F_{2b a_2\cdots a_n}
p_3^b\right)\,\,.\nonumber\eeqa Note that all $D$-dependence and
$m$-dependence cancel out. Now comparing this field theory
amplitude with the string theory amplitude \reef{expan1} in which
$\alpha$ is given by \reef{alpha}, one finds exact agreement if
normalizes the string theory amplitude \reef{a2} by factor
$-i/(64\pi)$.

The $t$-channel amplitude in field theory is given by the
following Feynman rule:\beqa A'_t&=&(\hV_{F_1T_4\bC})^{a_1\cdots
a_{n-1}} (\hG_{\bC})_{a_1\cdots a_{n-1}}{}^{b_1\cdots
b_{n-1}}(\hV_{\bC F_2T_3})_{b_1\cdots
b_{n-1}}\,\,,\labell{a8}\eeqa where the propagator and the vertex
function are \beqa (\hG_{\bC})^{a_1\cdots a_{n-1}}{}_{b_1\cdots
b_{n-1}} &=&\frac{i(n-1)!}{t}\eta^{[a_1}_{[b_1}\eta^{a_2}_{b_2}
\cdots\eta^{a_{n-1}]}_{b_{n-1}]}\,\,,\nonumber\\
(\hV_{T_3F_2\bC})^{a_1\cdots
a_{n-1}}&=&-\frac{a_T}{(n-1)!}p_{3a}F_2^{a a_1\cdots
a_{n-1}}\,\,.\nonumber\eeqa Replacing them in the amplitude
\reef{a8}, one finds \beqa
A'_t&=&\frac{-ia_T^2}{(n-1)!t}p_{3a}F_2^{a a_1\cdots
a_{n-1}}F_1^b{}_{a_1\cdots a_{n-1}}p_{4b}\,\,.\nonumber\eeqa For
simplicity, consider only the terms that have
$\veps_1\inn\veps_2$. Simple algebra reduces above amplitude to
the following: \beqa A_t'&=&-\frac{ia_T^2}{2(n-1)!t}(-p_1\inn
p_2p_3\inn p_4+p_3\inn p_2 p_4\inn p_1+p_4\inn p_2 p_3\inn
p_1)\veps_1^{a_1\cdots a_{n-1}}\veps_{2 a_1\cdots
a_{n-1}}\labell{at}\\
&&-\frac{ia_T^2}{2(n-1)!}p_1\inn p_2 \veps_1^{a_1\cdots
a_{n-1}}\veps_{2 a_1\cdots a_{n-1}}+\cdots\,\,.\nonumber\eeqa
Similarly, the $u$-channel in field theory is \beqa
A_u'&=&\frac{-ia_T^2}{2(n-1)!u}(-p_1\inn p_2p_3\inn p_4+p_3\inn
p_2 p_4\inn p_1+p_4\inn p_2 p_3\inn p_1)\veps_1^{a_1\cdots
a_{n-1}}\veps_{2 a_1\cdots
a_{n-1}}\labell{au}\\
&&-\frac{ia_T^2}{2(n-1)!}p_1\inn p_2 \veps_1^{a_1\cdots
a_{n-1}}\veps_{2 a_1\cdots a_{n-1}}+\cdots\,\,.\nonumber\eeqa
Comparing the above poles with the corresponding poles in string
theory \reef{expan1}, one finds \beqa
a_T^2&=&1/2\,\,.\labell{ab}\eeqa Moreover, imposing the fact the
string theory amplitude does not have the above contact terms
fixes the constant $b$  to be \beqa
b_T&=&a_T^2/2\,=\,1/4\,\,.\labell{abc}\eeqa The next order terms
in \reef{expan1} are related to eight derivative order terms in
the action in which we are not interested in the present paper.

For the scalar action, one may add the following couplings to the
action \reef{action00}: \beqa S_2^g&=&-\int
d^{D}x\sqrt{G}\left[\frac{1}{2}(F_{(n)}\inn
F_{(n)}+\bF_{(n)}\inn\bF_{(n)})(1+a_gg+b_gg^2)\right]\,\,,\labell{action1}\eeqa
The $s$-channel amplitude is exactly like the amplitude \reef{a6}
in which tachyons are replaced by the scalars. Accordingly, one
finds exact agreement with the first term in string amplitude
\reef{a5}. The $t$- and $u$-channel again are like in the tachyon
case in which the tachyons are replaced by the scalar, and $\bC$
is replaced by $C$, hence, one finds the result \reef{at} and
\reef{au}, respectively. Comparing them with string theory
amplitude \reef{a5}, one finds \beqa
a_g\,=\,1/4\,\,\,\,&;&\,\,\,\,b_g\,=\,a_g^2/2\,=\,1/8\labell{abc1}\eeqa

\section{Discussion}

In this paper, we have evaluated various  sphere level S-matrix
elements involving tachyon vertex operators in type 0 theory. We
then find an expansion for these amplitudes that their leading
order terms are correspond to covariant tachyon action. The two
derivatives order action that we have found, \reef{action0},
\reef{action10}, \reef{ab}, and \reef{abc}, in terms of
$F^{\pm}_{(n)}=(F_{(n)}\pm\bF_{(n)})/\sqrt(2)$, is \beqa
S^T&=&-\int
d^Dx\sqrt{G}\left[e^{-2\Phi}\left(-2R-8\prt^a\Phi\prt_a\Phi+\frac{3}{2}H^2+
\frac{1}{2}
\prt^aT\prt_aT+\frac{1}{2}m^2T^2\right)\right.\nonumber\\
&&\left.+\frac{1}{2}\left(F^{+}_{(n)}\inn
F^{+}_{(n)}\right)f(T)+\frac{1}{2}\left(F^{-}_{(n)}\inn
F^{-}_{(n)}\right)f(-T)\right]+\cdots\,\,,\labell{action2}\eeqa
where dots represent couplings that are of order eight
derivatives and higher, and their coefficients include
$\z(3),\,\z(4),\cdots$. The function $f(T)$ is \beqa
f(T)&=&1\pm\frac{1}{\sqrt{2}}T+\frac{1}{4}T^2+\cdots\,\,.\nonumber\eeqa

The tachyon coupling $F\bF T$ that we have extracted from the
S-matrix element of two RR and two tachyons, can also be
extracted from the S-matrix element of four RR states with
opposite chirality. The coupling $F\bF T$ appears in this
amplitude as a tachyonic pole. However, this amplitude can fix
the sum of $F\bF T$ and $FF\bF\bF$ couplings. Since we don't know
the coupling $FF\bF\bF$, this study can not fix the coupling
$F\bF T$ without ambiguity. We analyze to some extent this
S-matrix element in the Appendix B.

The action for the scalar field that we have found,
\reef{action00}, \reef{action1} and \reef{abc1},  is the
following: \beqa S^g&=&-\int
d^Dx\sqrt{G}\left[e^{-2\Phi}\left(-2R-8\prt^a\Phi\prt_a\Phi+\frac{3}{2}H^2+
\frac{1}{2}
\prt^ag\prt_ag\right)\right.\nonumber\\
&&\left.+\frac{1}{2}\left(F_{(n)}\inn
\bF_{(n)}\right)\left(1\pm\frac{1}{2}g+\frac{1}{8}g^2+\cdots\right)
\right]\,\,,\labell{action22}\eeqa This action is the low energy
action for the scalar field. This action should be consistent
with the dimensional reduction of the following 10-dimensional
action: \beqa S&\!\!=\!\!&-\int
d^{10}x\sqrt{G}\left[e^{-2\Phi}\left(-2R-8\prt^a\Phi\prt_a\Phi+\frac{3}{2}H^2\right)
+\frac{1}{2}\left(F_{(n)}\inn
F_{(n)}\right)+\frac{1}{2}\left(\bF_{(n)}\inn
\bF_{(n)}\right)\right]\,\,.\nonumber\eeqa Dimensional reduction
of this  action to 9-dimension is \cite{ebrk} \beqa S&=&-\int d^9x
\sqrt{G}\left[e^{-2\Phi}\left(-2R-8\prt^a\Phi\prt_a\Phi+\frac{3}{2}H^2+
2\prt^a(\log k)\prt_a(\log k)\right)\right.\nonumber\\
&&\left.+\frac{1}{2}\left(F_{(n)}\inn F_{(n)}\right)k+\frac{1}{2}
\left(\bF_{(n)}\inn \bF_{(n)}\right)k\right]\,\,, \nonumber\eeqa
where the scalar $k$ is related to the $G_{1010}$ component of the
metric \cite{ebrk}. Note that we have considered   only those
Kalb-Ramond and RR fields that have components in the non-compact
space, and only $G_{ab}$ and $G_{1010}$ component of metric.
These are the fields  that we have studied in our paper. Using
the field redefinition    $g= \pm 2\log k$ to write the kinetic
term in the standard form, one finds
 \beqa S&=&-\int d^9x
\sqrt{G}\left[e^{-2\Phi}\left(-2R-8\prt^a\Phi\prt_a\Phi+\frac{3}{2}H^2+
\frac{1}{2}\prt^a g\prt_a g\right)\right.\nonumber\\
&&\left.+\frac{1}{2}\left(F_{(n)}\inn F_{(n)}\right)e^{\pm
g/2}+\frac{1}{2} \left(\bF_{(n)}\inn \bF_{(n)}\right)e^{\pm
g/2}\right]\,\,.\labell{action5} \eeqa It is easy to see that the
action \reef{action22} is consistent with above action as
expected. The above action is also consistent with the fact that,
due to the vanishing of their  world-sheet  correlation functions,
self-coupling of odd number of scalars and coupling of odd number
of scalars to graviton are zero.

We have been assuming  throughout the section 2.1 that there is no
tachyon coupling $RT^2$. This assumption was also made in
\cite{ikat}. This is consistent with the observation made in
section 2 that  the S-matrix element of two tachyons and two
gravitons,  and the S-matrix element of two scalars and two
gravitons can be written in the identical form \reef{a166}. This
indicates that  apart from those tachyon couplings that the mass
$m$ appears as their coefficient, the coupling of two tachyons
and two gravitons, and the coupling of two scalars and two
gravitons should be the same. Since the scalar action
\reef{action5} has no coupling $Rg^2$, accordingly, the tachyon
action has no coupling $RT^2$ either. Similarly, following the
discussion in the last paragraph in section 3.1, on concludes
that there is no tachyon coupling $T^2\prt_aT\prt^aT$ because
there is no $g^2\prt_ag\prt^ag$ coupling in the scalar action
\reef{action5}. This is unlike the open string tachyon case that
such a coupling is non-zero \cite{mg,mrga}.

We have seen that our calculation fixes the form of tachyon
potential to be \reef{a21}.  This potential has no local minimum.
This means the stability of theory can not be reached through
condensation of only tachyon field. As pointed out in
\cite{ikat}, however, the instability may be cured in the
presence of background RR field. In the presence of background
flux $F$ and $\bF$, the on-shell tachyon potential, up to
quadratic order of background field, is the following \beqa
V(T)&=&-\frac{1}{2}T^2+\frac{1}{2}F^2f(T)+\frac{1}{2}\bF^2f(-T)\,\,.\nonumber\eeqa
For appropriate background flux,  it may have local minimum that
the unstable  theory at $T=0$ condenses to it. In principle,
however, one may use a field redefinition such that in the new
variables the instability would be cured by condensation of only
tachyon. In this regard, it was found in \cite{oa} that the
tachyon potential in the sigma model approach which is expected
to be related to S-matrix based approach by field redefinition,
has local minimum.

We have seen that the tachyon and scalar have, to the leading
order, similar couplings in the bulk. Now it raises a question:
does the tachyon and massless scalars have also similar couplings
to D-branes? To answer this question,  consider the coupling of
two tachyons to D-branes of type 0 theory. This amplitude is the
following \cite{mrga1}: \beqa
A(T,T)&\sim&\frac{\Ga(-t/2)\Ga(-2s)}{\Ga(-1-t/2-2s)}\,\,,\labell{a20}\eeqa
where $t=-\alpha'(p_1+p_2)^2/2$ and $s=-\alpha'(p_1 \inn G \inn
p_1)/2$ where $G$ here stands for the open string metric. Now
compare it with the scattering amplitude of two massless scalars
from D-brane. This amplitude can be read from the general result
in  \cite{mg6}, \beqa
A(g,g)&\sim&\frac{\Ga(-t/2)\Ga(-2s)}{\Ga(1-t/2-2s)}
\left(\frac{}{}4s^2+ t(s+t/4)\right)\,\,,\nonumber\eeqa where we
have assumed there is only one scalar, \ie $\z\rightarrow 1$. Note
that the Mandelstam variables in these amplitudes are arbitrary.
As it can be seen,  they are not identical amplitude. That means
the coupling of tachyon and the scalars to D-branes are not
similar. In other words, if one expands both in the limit
$s,t\rightarrow 0$, one finds \beqa
A(T,T)&\sim&2\left(\frac{1+2s}{t}-\frac{p_{1}\inn N\inn
p_2}{4s}+\frac{3}{4}
+\cdots\right)\,\,,\nonumber\\
A(g,g)&\sim&2\left(\frac{2s}{t}-\frac{p_{1}\inn N\inn
p_2}{4s}+\frac{1}{4} +\cdots\right)\,\,,\nonumber\eeqa where  $N$
is the flat metric in the space orthogonal to the D-brane. The
massless poles are reproduce by standard  covariant action
\cite{mrga1,ikat1}. The contact terms above indicates that the
quadratic tachyon coupling to D-branes is different from the
quadratic scalar coupling to the D-branes.

Finally, we note that the expansion of the S-matrix elements of
massless scalar states \reef{a16666} and \reef{a10} in the limit
that the Mandelstam variables approach zero has, in general,
undesirable Euler-Mascheroni constant $\ga=0.5772157$. When one
imposes the on-shell constraint on the Mandelstam variables they
disappear. However, these  S-matrix elements in the form
\reef{a166} and \reef{a12} have no such undesirable terms. They
already disappear by imposing the constrains in the amplitude.
For example, consider on-shell S-matrix element of four scalars
\reef{a10}. The first term in this equation has the following
expansion: \beqa
\frac{\Ga(1-\frac{u}{2})\Ga(-\frac{s}{2})\Ga(1-\frac{t}{2})}
{\Ga(\frac{u}{2})\Ga(1+\frac{s}{2})\Ga(\frac{t}{2})}&=&-\frac{ut}{2s}-
\frac{\ga
u(s+t+u)t}{2s}-\frac{\ga^2u(s+t+u)^2t}{4s}+\cdots\,\,,\nonumber\eeqa
if one imposes the on-shell constraint $s+t+u=0$, the terms that
have $\ga$ vanishes. Now the same term in the  S-matrix element
in the form \reef{a12} has the following expansion: \beqa
\frac{\Ga(1+\frac{s+t-u}{4})\Ga(-\frac{s}{2})\Ga(1+\frac{s+u-t}{4})}
{\Ga(\frac{u-s-t}{4})\Ga(1+\frac{s}{2})\Ga(\frac{t-u-s}{4})}&=&-
\frac{s^2-(t-u)^2}{8s}-\frac{\z(3)}{128}(s^2-(t-u)^2)^2+\cdots\,\,.\nonumber\eeqa
The constant $\ga$ does not appear in the amplitude at all. This
may indicate that the constraint is imposed in the amplitude
correctly.


{\bf Acknowledgement}: I would like to thank A. A. Tseytlin for
comments.
\newpage
{\large {\bf Appendix A}}

In this appendix we give the result for some integrals that
appear in the previous sections. Consider the following integral:
\beqa I&=&\int
d^2z\,|z|^{2a}|1-z|^{2b}f(z,\bz)\,\,.\labell{I}\eeqa To evaluate
this integral, one should write \beqa
|z|^{2a}&=&\frac{1}{\Ga(-a)}\int_0^{\infty}dt
t^{-a-1}e^{-t|z|^2}\,\,,\nonumber\eeqa and similarly for
$|1-z|^{2b}$. This turns the $z$ integration into an elementary
integration that can be explicitly carried out. In this way \beqa
I&=&\frac{1}{\Ga(-a)\Ga(-b)}\int_0^{\infty}
dtdu\,t^{-a-1}u^{-b-1}J(t,u)\,\,,\nonumber\eeqa where \beqa
J(t,u)&=&\int
d^2z\,e^{-t|z|^2-u|1-z|^2}f(z,\bz)\,\,.\nonumber\eeqa This
integral is easy to evaluate for some simple function $f(z,\bz)$.
After performing this integral, one should make the change of
variables $t=x/s$ and $u=x/(1-s)$. Then using the following
definitions for gamma and beta functions: \beqa
\Ga(\alpha)&=&\int_0^{\infty}dx\,x^{\alpha-1}e^{-x}\,\,,\nonumber\\
B(\alpha,\beta)&=&\int_0^1ds\,(1-s)^{\alpha}s^{\beta}\,\,,\nonumber\eeqa
one finds the final result for the integral \reef{I} in terms of
gamma functions. For example \beqa \int
d^2z\,|z|^{2a}|1-z|^{2b}&=&2\pi\frac{\Ga(a+1)\Ga(b+1)
\Ga(-a-b-1)}{\Ga(-a)\Ga(-b)\Ga(a+b+2)}\,\,,\nonumber\\
\int d^2z\,|z|^{2a}|1-z|^{2b}z&=&2\pi\frac{\Ga(a+2)\Ga(b+1)
\Ga(-a-b-1)}{\Ga(-a)\Ga(-b)\Ga(a+b+3)}\,\,,\nonumber\\
\int d^2z\,|z|^{2a}|1-z|^{2b}\bz&=&2\pi\frac{\Ga(a+2)\Ga(b+1)
\Ga(-a-b-1)}{\Ga(-a)\Ga(-b)\Ga(a+b+3)}\,\,,\nonumber\\
\int d^2z\,|z|^{2a}|1-z|^{2b}(1-z)&=&2\pi\frac{\Ga(a+1)\Ga(b+2)
\Ga(-a-b-1)}{\Ga(-a)\Ga(-b)\Ga(a+b+3)}\,\,,\nonumber\\
\int d^2z\,|z|^{2a}|1-z|^{2b}(1-\bz)&=&2\pi\frac{\Ga(a+1)\Ga(b+2)
\Ga(-a-b-1)}{\Ga(-a)\Ga(-b)\Ga(a+b+3)}\,\,,\nonumber\\
\int
d^2z\,|z|^{2a}|1-z|^{2b}(1-\bz)z&=&-2\pi\frac{\Ga(a+2)\Ga(b+2)
\Ga(-a-b-2)}{\Ga(-a)\Ga(-b)\Ga(a+b+3)}\,\,,\nonumber\\
\int
d^2z\,|z|^{2a}|1-z|^{2b}(1-z)\bz&=&-2\pi\frac{\Ga(a+2)\Ga(b+2)
\Ga(-a-b-2)}{\Ga(-a)\Ga(-b)\Ga(a+b+3)}\,\,,\nonumber\\
\int d^2z\,|z|^{2a}|1-z|^{2b}z^2&=&2\pi\frac{\Ga(a+3)\Ga(b+1)
\Ga(-a-b-1)}{\Ga(-a)\Ga(-b)\Ga(a+b+4)}\,\,,\nonumber\\
\int d^2z\,|z|^{2a}|1-z|^{2b}\bz^2&=&2\pi\frac{\Ga(a+3)\Ga(b+1)
\Ga(-a-b-1)}{\Ga(-a)\Ga(-b)\Ga(a+b+4)}\,\,,\nonumber\\
\int
d^2z\,|z|^{2a}|1-z|^{2b}(1-\bz)^2&=&2\pi\frac{\Ga(a+1)\Ga(b+3)
\Ga(-a-b-1)}{\Ga(-a)\Ga(-b)\Ga(a+b+4)}\,\,,\nonumber\\
\int d^2z\,|z|^{2a}|1-z|^{2b}(1-z)^2&=&2\pi\frac{\Ga(a+1)\Ga(b+3)
\Ga(-a-b-1)}{\Ga(-a)\Ga(-b)\Ga(a+b+4)}\,\,,\nonumber\\
\int
d^2z\,|z|^{2a}|1-z|^{2b}(1-\bz)^2z^2&=&2\pi\frac{\Ga(a+3)\Ga(b+3)
\Ga(-a-b-3)}{\Ga(-a)\Ga(-b)\Ga(a+b+4)}\,\,,\nonumber\\
\int
d^2z\,|z|^{2a}|1-z|^{2b}(1-z)^2\bz^2&=&2\pi\frac{\Ga(a+3)\Ga(b+3)
\Ga(-a-b-3)}{\Ga(-a)\Ga(-b)\Ga(a+b+4)}\,\,.\nonumber \eeqa
\newpage
{\large {\bf Appendix B}}

In this appendix, we evaluate the S-matrix element of four RR
with opposite chirality, and then compare it with field theory.
In world-sheet conformal field theory, the S-matrix element of two
$CC$ and two $\bC\bC$ states is given by the following
correlation function: \beqa A&\sim& \langle:V_{(-1/2,-1/2)}^{\rm
RR}(p_1,\veps_1):V_{(-1/2,-1/2)}^{\rm
RR}(p_2,\veps_2):V_{(-1/2,-1/2)}^{\rm
RR}(p_3,\veps_3):V_{(-1/2,-1/2)}^{\rm RR}
(p_4,\veps_4):\rangle\,\,,\labell{amp4}\nonumber\eeqa where the RR
vertex operators are given in \reef{ver3}. Two of them has
positive (negative) chirality and two others negative (positive)
chirality. The nontrivial correlation is the correlation between
the four spin operator with opposite chirality. It is give by the
following relation \cite{vk}: \beqa
<:S_A(z_1):S_B(z_2):S_{\dot{C}}(z_3):S_{\dot{D}}:>&=&\frac{1}{2}
(\ga_{\mu})_{AB}(\ga^{\mu})_{\dot{C}\dot{D}}(z_{13}z_{14}z_{23}
z_{24})^{-1/4}(z_{12}z_{34})^{-3/4}\nonumber\\
&&+{\cal C}_{A\dot{C}}{\cal C}_{B\dot{D}}(z_{12}z_{34})^{1/4}
(z_{14}z_{23})^{-1/4}(z_{13}z_{24})^{-5/4}\nonumber\\
&&-{\cal C}_{A\dot{D}}{\cal C}_{B\dot{C}}(z_{12}z_{34})^{1/4}
(z_{13}z_{24})^{-1/4}(z_{14}z_{23})^{-5/4}\,\,,\nonumber\eeqa
where ${\cal C}$ is the charge conjugation matrix. The other
correlators are easy to evaluate. The final result, after fixing
its $SL(2,R)$
 symmetry and doing some algebra on the gamma matrices, is
 \beqa A&\sim&
2\pi\left[\frac{1}{8}\Tr(P_{\mp}\Ga_{1(n)}\ga_{\nu}\Ga_{2(n)}\ga_{\mu})
\Tr(P_{\pm}\Ga_{3(n)}\ga^{\nu}\Ga_{4(n)}\ga^{\mu})\frac{\Ga(\frac{1}{2}-\frac{u}{2})
\Ga(-\frac{s}{2})\Ga(\frac{1}{2}-\frac{t}{2})}{\Ga(\frac{1}{2}+\frac{u}{2})
\Ga(1+\frac{s}{2})\Ga(\frac{1}{2}+\frac{t}{2})}\right.\nonumber\\
&&\left.+\Tr(P_{\pm}\Ga_{3(n)}\Ga_{1(n)})\Tr(P_{\pm}\Ga_{4(n)}\Ga_{2(n)})
\frac{\Ga(-\frac{1}{2}-\frac{u}{2})
\Ga(1-\frac{s}{2})\Ga(\frac{1}{2}-\frac{t}{2})}{\Ga(\frac{3}{2}+\frac{u}{2})
\Ga(1+\frac{s}{2})\Ga(\frac{1}{2}+\frac{t}{2})}\right.\nonumber\\
&&\left.+\frac{1}{2}\left(\Tr(P_{\pm}\ga_{\mu}\Ga_{1(n)}\Ga_{3(n)}
\ga^{\mu}\Ga_{4(n)}\Ga_{2(n)})
+\Tr(P_{\pm}\Ga_{2(n)}\ga_{\mu}\Ga_{1(n)}\Ga_{3(n)}\ga^{\mu}\Ga_{4(n)})\right)\right.
\nonumber\\
&&\left.\times \frac{\Ga(\frac{1}{2}-\frac{u}{2})
\Ga(1-\frac{s}{2})\Ga(\frac{1}{2}-\frac{t}{2})}{\Ga(\frac{3}{2}+\frac{u}{2})
\Ga(1+\frac{s}{2})\Ga(\frac{1}{2}+\frac{t}{2})}\right.\labell{a33}\\
&&\left.-\Tr(P_{\pm}\Ga_{3(n)}\Ga_{2(n)}\Ga_{4(n)}\Ga_{1(n)})
 \frac{\Ga(\frac{1}{2}-\frac{u}{2})
\Ga(1-\frac{s}{2})\Ga(\frac{1}{2}-\frac{t}{2})}{\Ga(\frac{3}{2}+\frac{u}{2})
\Ga(1+\frac{s}{2})\Ga(\frac{3}{2}+\frac{t}{2})} +3\leftrightarrow
4 \right]\,\,,\nonumber\eeqa where the Mandelstam variables are
given in \reef{mandel}. Massless pole appears only in the  first
term, and all other terms have  tachyonic or massive poles. They
contribute to contact terms in the covariant limit/expansion, \ie
the expansion at $s,t,u\rightarrow 0$..

To normalize the amplitude, we consider, for simplicity, $n=1$.
The momentum expansion  is \beqa
A&\sim&2\pi(32)^2\left[-\frac{1}{16s}\left(\frac{}{}2p_1\inn
p_3p_2\inn p_4+(D-4)p_1\inn p_2 p_3\inn p_4+2p_1\inn p_4 p_2\inn
p_3\right)+\cdots +3\leftrightarrow
4\right]\,\,,\labell{a34}\nonumber\eeqa where dots represent
contact terms that have at least four momenta. In field theory,
the massless pole is given by the following Feynman rule: \beqa
A_s'&=&(\hV_{F_1F_2h})_{ab}(\hG_h)^{ab}{}_{cd}(\hV_{h\bF_3\bF_4})^{cd}+
\hV_{F_1F_2\Phi'}\hG_{\Phi'}V_{\Phi'\bF_3\bF_4}\nonumber\\
&=& -\frac{i}{4s}\left[(D-4)p_1\inn p_2p_3\inn p_4+2p_1\inn
p_3p_4\inn p_2+2p_1\inn p_4 p_3\inn p_2\right]\,\,,\nonumber\eeqa
which is exactly the massless pole of string theory provided one
normalizes the amplitude \reef{a33} by factor $i/(\pi(32)^2)$.
The next order terms  correspond to two different terms in field
theory. One is the contact term $FF\bF\bF$, and the other the
tachyonic pole resulting from two $F\bF T$ couplings and tachyon
as propagator. The propagator should be replaced by 1. Hence, the
next order terms can fix the sum of $FF\bF\bF$ and $F\bF T$
couplings.


\begin{thebibliography}{99}
\bibitem
{anjs}{A. Neveu and J. Scherk, Nucl. Phus. B {\bf 36}, 155
(1972);\\
J. Scherk and J. Schwartz, Nucl. Phys. B {\bf 81}, 118 (1974).}
\bibitem
{mgas}{M. Gutperle and A. Strominger,  JHEP {\bf 0204},018 (2002)
[arXiv: hep-th/0202210].}
\bibitem
{asen1}{A. Sen,  JHEP {\bf 0204}, 048 (2002) [arXiv:
hep-th/0203211].}
\bibitem
{asen2}{A. Sen,   JHEP {\bf 0207}, 065 (2002) [arXiv:
hep-th/0203265].}
\bibitem
{asen3}{A. Sen,  Mod. Phys. Lett. A {\bf 17}, 1797 (2002) [arXiv:
hep-th/0204143].}
\bibitem
{asen4}{A. Sen,  JHEP {\bf 0210}, 003 (2002) [arXiv:
hep-th/0207105].}
\bibitem
{asen5}{A. Sen, ``Time and tachyon,'' arXiv: hep-th/0209122.}
\bibitem
{pmas}{P. Mukhopadhyay and A. sen,  JHEP {\bf 0211}, 047 (2002)
[arXiv: hep-th/0208142].}
\bibitem
{as}{A. Strominger, ``Open string creation by S-branes,'' arXiv:
hep-th/0209090.}
\bibitem
{flanst}{F. Larsen, A. Naqvi and S. Terashima,  JHEP {\bf 0302},
039 (2003) [arXiv: hep-th/0212248].}
\bibitem
{mgas1}{M. Gutperle and A. Strominger, Phys. Rev. D {\bf 67},
126002 (2003) [arXiv: hep-th/0301038].}
\bibitem
{amas}{A. Maloney, A. Strominger and X. Yin, ``S-brane
thermodynamica,'' arXiv: hep-th/0302146.}
\bibitem
{toss}{T. Okuda and S. Sugimoto,  Nucl. Phys. B {\bf 647}, 101
(2002) [arXiv: hep-th/0208196].}
\bibitem
{srss}{S. J. Rey and S. Sugimoto, Phys. Rev. D {\bf 67}, 086008
(2003) [arXiv: hep-th/0301049].}
\bibitem
{srss1}{S. J. Rey and S. Sugimoto, Phys. Rev. D {\bf 68}, 026003
(2003)  [arXiv: hep-th/0303133].}
\bibitem
{nrc}{N.R. Constable and F. Larsen, JHEP {\bf 0306}, 017 (2003)
[arXiv: hep-th/0305177].}
\bibitem
{bcml}{B. Chen, M. Li and F.L. Lin, JHEP {\bf 0211}, 050 (2002)
[arXiv: hep-th/0209222].}
\bibitem
{ss}{S. Sugimoto and S. Terashima, JHEP {\bf 0207}, 025 (2002)
[arXiv: hep-th/0205085].}
\bibitem
{jam}{J.A. Minahan, JHEP {\bf 0207}, 030 (2002) [arXiv:
hep-th/0205098].}
\bibitem
{nm}{N. Moeller and B. Zwiebach, JHEP {\bf 0210}, 034 (2002)
[arXiv: hep-th/0207107].}
\bibitem
{mf}{M. Fujita and H. Hata, JHEP {\bf 0305}, 043 (2003) [arXiv:
hep-th/0304163].}
\bibitem
{iya}{I.Y. Aref'eva, L.V. Joukovskaya and A.S. Koshelev, ``Time
Evolution in Superstring Field Theory on non-BPS brane.I. Rolling
Tachyon and Energy-Momentum Conservation,'' arXiv:
hep-th/0301137.}
\bibitem
{nlhl}{N. Lambert, H. Liu and J. Maldacena, ``Closed strings from
decaying D-branes,'' [arXiv: hep-th/0303139].}
\bibitem
{jm}{J. McGreevy and Verlinde, ``Strings from Tachyons,'' arXiv:
hep-th/0304224.}
\bibitem
{jm2}{J. McGreevy, J. Teschner and H. Verlinde, ``Classical and
Quantum D-branes in 2D String Theory,'' arXiv: hep-th/0305194.}
\bibitem
{irk}{I.R. Klebanov, J. Maldacena and N. Seiberg, ``D-brane Decay
in Two-Dimensional String Theory,'' arXiv: hep-th/0305159.}
\bibitem
{sen1}{A. Sen, ``Open and Closed Strings fron Unstable
D-branes,'' arXiv: hep-th/0305011.}
\bibitem
{mrd}{M.R. Douglas, I.R. Klebanov, D. Kutasov, J. Maldacena, E.
Martinec and N. Seiberg, ``A New Hat For The C=1 Matrix Model,''
arXiv:hep-th/0307195.}
\bibitem
{jlk}{J.L. Karczmarek, H. Liu, J. Maldacena and A. Strominger,
``UV Finite Brane Decay,'' arXiv: hep-th/0306132.}
\bibitem
{sen2}{A. Sen, ``Open-Closed Duality at Tree Level,'' arXiv:
hep-th/0306137.}
\bibitem
{aatt}{A.A. Tseytlin, J. Math. Phys. {\bf 42}, 2854 (2001)
[arXiv: hep-th/0011033].}
\bibitem
{ku}{D. Kutasov, V. Niarchos, Nucl. Phys. B {\bf 666}, 56 (2003)
[arXiv: hep-th/0304045].}
\bibitem
{ms}{ M. Smedb${\rm \ddot{a}}$ck, ``On Effective Actions for the
Bosonic
Tachyon,'' hep-th/0310138;\\
A. Fotopoulos, A.A. Tseytlin, ``On open superstring partition
function in inhomogeneous rolling tachyon background,''
hep-th/0310253.}
\bibitem
{wt}{E. Coletti, I. Sigalov and W. Taylor, ``Abelian and
non-abelian vector field effective actions from string field
theory,'' arXiv: hep-th/0306041.}
\bibitem
{mrga}{M.R. Garousi, ``S-matrix elements and off-shell tachyon
action with non-abelian gauge symmetry,'' arXiv: hep-th/0307197.}
\bibitem
{oa}{O. Andreev, ``Comments on Tachyon Potentials in Closed and
Open-Closed String Theories,'' arXiv: hep-th/0308123.}
\bibitem
{mg}{M. R. Garousi, Nucl. Phys. B {\bf 584}, 284 (2000)
[arXiv:hep-th/0003122];  Nucl. Phys. B {\bf 647}, 117 (2002)
[arXiv: hep-th/0209068];  JHEP {\bf 0304}, 027 (2003) [arXiv:
hep-th/0303239]; JHEP {\bf 0305}, 058 (2003) [arXiv:
hep-th/0304145].}
\bibitem
{ebmr}{E. A. Bergshoeff, M. de Roo, T. C. de Wit, E. Eyras and S.
Panda, JHEP {\bf 0005}, 009 (2000) [arXiv:hep-th/0003221];\\
J. Kluson, Phys. Rev. D {\bf 62}, 126003 (2000)
[arXiv:hep-th/0004106].}
\bibitem
{ldjh}{L.J. Dixon and J.A. Harvey, Nucl. Phys. B {\bf 274}, 93
(1986).}
\bibitem
{mg6}{M. R. Garousi and R. C. Myers,  Nucl. Phys. B {\bf 475},
193 (1996) [arXiv:hep-th/9603194].}
\bibitem
{jp}{J. Polchinski, ``Superstring Theory,'' Cambridge University
Pree (1998).}
\bibitem
{ikat}{I.R. Klebanov and A.A. Tseytlin, Nucl. Phys. B {\bf 546},
155 (1999) [arXiv:hep-th/9811035].}
\bibitem
{aatt2}{A.A. Tseytlin, ``Type 0 strings and gauge theory,''
Imperial-th-99-00-02, 1999. 5pp. Contributed to 14th
International Workshop on High Energy Physics and Quantum Field
Theory (QFTHEP 99), Moscow, Russia, 27 May - 2 Jun 1999.
Published in *Moscow 1999, High energy physics and quantum field
theory* 427-430.}
\bibitem
{ebrk}{E. Bergshoeff, R. Kallosh and T. Ortin, Phys. Rev. D {\bf
51}, 3009 (1995) [arXiv:hep-th/9410230]; \\
E. Bergshoeff, C. Hull and T. Ortin, Nucl. Phys. B {\bf 451}, 547
(1995) [arXiv:hep-th/9504081].}
\bibitem
{pmto}{P. Meessen and T. Ortin, Phys. Rev. D {\bf 64}, 126005
(2001) [arXiv:hep-th/0103244].}
\bibitem
{mrga1}{M.R. Garousi, Nucl. Phys. B {\bf 550}, 225 (1999)
[arXiv:hep-th/9901085].}
\bibitem
{ikat1}{I.R. Klebanov and A.A. Tseytlin, Nucl. Phys. B {\bf 547},
143 (1999) [arXiv:hep-th/9812089].}
\bibitem
{vk}{V.A. Kostelecky, O. Lechtenfeld, W. Lerche, S. Samuel and S.
Watamura, Nucl. Phys. B {\bf 288}, 173 (1987).}
\end{thebibliography}
\end{document}